\documentclass[sigconf]{acmart}
\usepackage{acmart-taps}
\usepackage{makecell}
\usepackage{multirow}
\usepackage{graphicx}% http://ctan.org/pkg/graphicx
\usepackage{array}
\usepackage{color}
\usepackage{stfloats}
\usepackage{colortbl}

\newenvironment{revisedenv}{}{}%\color{revisedcolor}

\newcommand{\ipstart}[1]{\vspace{1mm} \noindent{\textbf{\textit{#1.}}}}

\AtBeginDocument{%
  \providecommand\BibTeX{{%
    \normalfont B\kern-0.5em{\scshape i\kern-0.25em b}\kern-0.8em\TeX}}}

\setcopyright{acmcopyright}
\copyrightyear{2023}
\acmYear{2023}
\setcopyright{rightsretained}
\acmConference[CHI '23]{Proceedings of the 2023 CHI Conference on Human Factors in Computing Systems}{April 23--28, 2023}{Hamburg, Germany}
\acmBooktitle{Proceedings of the 2023 CHI Conference on Human Factors in Computing Systems (CHI '23), April 23--28, 2023, Hamburg, Germany} 
\acmDOI{10.1145/3544548.3581494}
\acmISBN{978-1-4503-9421-5/23/04}

\newcommand{\sysname}{AVscript}
\newcommand{\andrew}{Bryan}
\newcommand{\rachel}{Rachel}
\newcommand{\lewis}{Lewis}

\begin{document}

%%
%% The "title" command has an optional parameter,
%% allowing the author to define a "short title" to be used in page headers.
\title{\sysname{}: Accessible Video Editing with Audio-Visual Scripts}
% \sysname{}: Text-based Video Editing using Audio-Visual Scripts for Screen-reader Users
% \sysname{}: Screen-Reader Accessible Video Editing using Audio-Visual Scripts
% \sysname{}: Audio-Visual Script for [Screen-Reader | Non-Visually] Accessible Video Editing
% \sysname{}: Editing Videos via Audio-Visual Script
% \sysname{}: Video Editing via Audio-Visual Scripts for Blind and Low Vision Video Creators
% \title{\sysname: Transcript-based Interface for Accessible Video Editing}
% terms: 
% text/script/transcript-based
% possibilities: screen-reader users, accessibility, blind and low vision, with visual impairments
% translating visuals to text, audio-visual script, audio-visual transcript, audio-visual text document
% 
% video editing (or video authoring)

% sysnames
% AVcut, AVEdit, 
% AVscript, scriptedit, scribedit
% ViSible
% VLens (lens = ?)
% ReadVid
% % Velvet (Video-Edit system for Low Vision Editors)
% AVenue (doesn't mean anything, AVid is good but used)
% Seethru
% Scriptify
% Scripto
% Viscript
% Scene reader (Screen reader like)
%CutScript ('Cut' is a symbolic action of movie creators like 'Final Cut', 'Director's Cut')
%A11iedCut

% \sysname{}: Text-Based Editing for Video Creators with Visual Impairments

%%
%% The "author" command and its associated commands are used to define
%% the authors and their affiliations.
%% Of note is the shared affiliation of the first two authors, and the
%% "authornote" and "authornotemark" commands
%% used to denote shared contribution to the research.
\author{Mina Huh}
\authornote{Mina Huh conducted part of this work as a research intern at NAVER AI Lab.}
\affiliation{
    \institution{The University of Texas at Austin}
    \city{Austin}
    \state{TX}
    \country{USA}
    }
\email{minahuh@cs.utexas.edu}

\author{Saelyne Yang}
\affiliation{
    \institution{KAIST}
    \city{Daejeon}
    \country{Republic of Korea}
    }
\email{saelyne@kaist.ac.kr}

\author{Yi-Hao Peng}
\affiliation{
    \institution{Carnegie Mellon University}
    \city{Pittsburgh}
    \state{PA}
    \country{USA}
    }
\email{yihaop@cs.cmu.edu}

\author{Xiang `Anthony' Chen}
\affiliation{
    \institution{UCLA}
    \city{Los Angeles}
    \state{CA}
    \country{USA}
    }
\email{xac@ucla.edu}

\author{Young-Ho Kim}
\affiliation{
    \institution{NAVER AI Lab}
    \country{Republic of Korea}
    }
\email{yghokim@younghokim.net}

\author{Amy Pavel}
\affiliation{
    \institution{The University of Texas at Austin}
    \city{Austin}
    \state{TX}
    \country{USA}
    }
\email{apavel@cs.utexas.edu}

%%
%% By default, the full list of authors will be used in the page
%% headers. Often, this list is too long, and will overlap
%% other information printed in the page headers. This command allows
%% the author to define a more concise list
%% of authors' names for this purpose.
%\renewcommand{\shortauthors}{Huh et al.} %Young-Ho: Our names are enough short to be listed on the header.

%%
%% The abstract is a short summary of the work to be presented in the
%% article.
\begin{abstract}
Sighted and blind and low vision (BLV) creators alike use videos to communicate with broad audiences. 
Yet, video editing remains inaccessible to BLV creators.
Our formative study revealed that current video editing tools make it difficult to access the visual content, assess the visual quality, and efficiently navigate the timeline.
We present ~\sysname{}, an accessible text-based video editor. ~\sysname{} enables users to edit their video using a script that embeds the video's visual content, visual errors (\textit{e.g.}, dark or blurred footage), and speech. Users can also efficiently navigate between scenes and visual errors or locate objects in the frame or spoken words of interest.
A comparison study (N=12) showed that ~\sysname{} significantly lowered BLV creators' mental demands while increasing confidence and independence in video editing. We further demonstrate the potential of ~\sysname{} through an exploratory study (N=3) where BLV creators edited their own footage.

\end{abstract}

%%
%% The code below is generated by the tool at http://dl.acm.org/ccs.cfm.
%% Please copy and paste the code instead of the example below.
%%

% \ccsdesc{Computer systems organization~Robotics}
% \ccsdesc[100]{Networks~Network reliability}

%%
%% Keywords. The author(s) should pick words that accurately describe
%% the work being presented. Separate the keywords with commas.

%% A "teaser" image appears between the author and affiliation
%% information and the body of the document, and typically spans the
%% page.
\begin{teaserfigure}
  \centering
  \includegraphics[width=\textwidth]{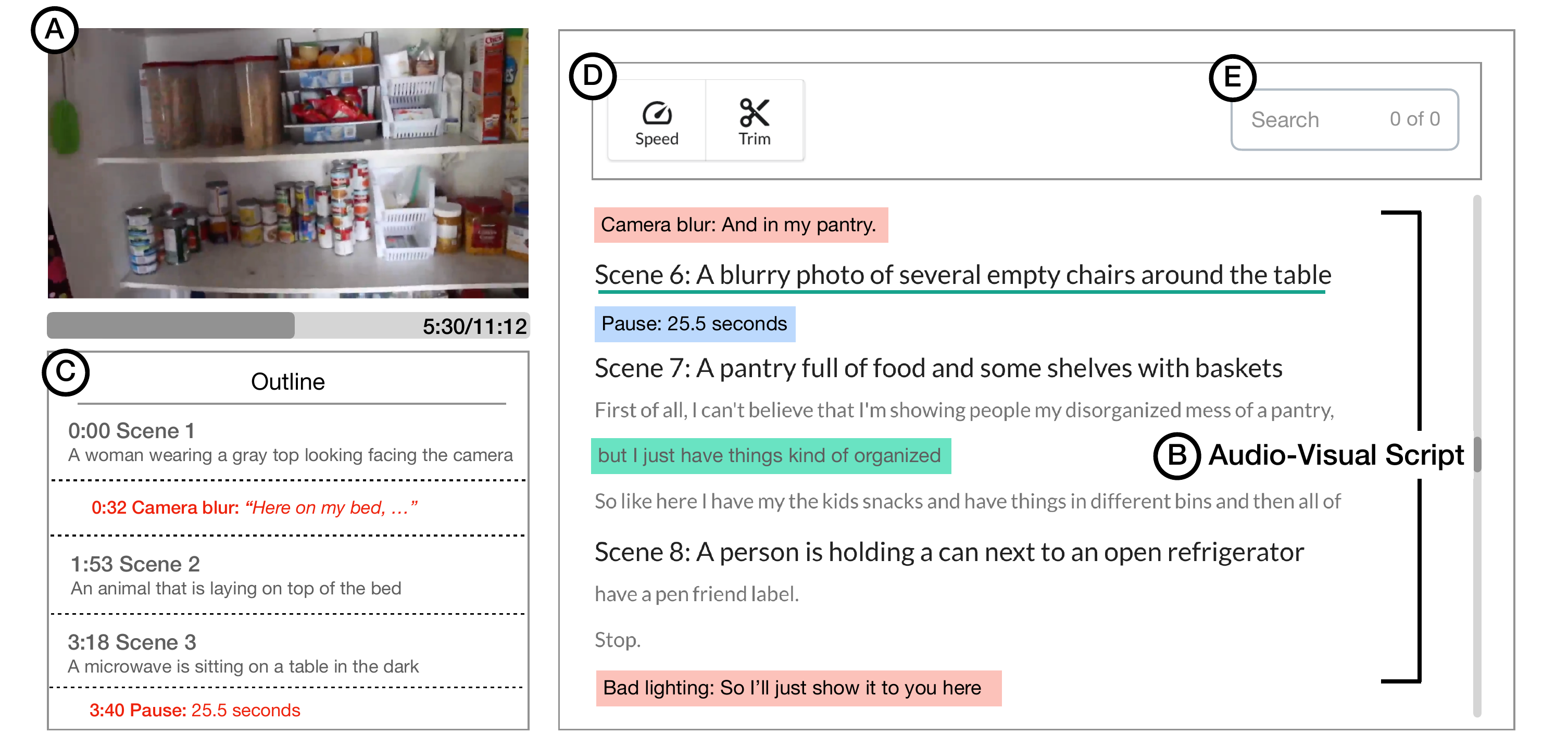}
  \caption{~\sysname{} is an accessible text-based video editing tool that enables blind and low-vision creators to edit videos efficiently using a screen reader. The ~\textit{video pane} (A) provides notifications for visual errors and supports inspection of visual objects. The ~\textit{audio-visual script} (B) provides a narration transcript segmented by scenes, scene descriptions, and highlighted visual errors (\textit{e.g.,} blur, bad lighting). The ~\textit{outline pane} (C) is a navigable summary of the video scenes and errors. The ~\textit{tool pane} (D) allows users to trim or change the playback speed of the selected video clip. The ~\textit{search pane} (E) supports both visual search and narration search of the video.}
  \label{fig:teaser}
    \Description{Teaser image of how \sysname{}'s components work. A video pane of \sysname{} is shown at the top left corner. There is a timeline bar indicating the progress of the video player and it shows that 5 minutes and 30 seconds has past for a 11 minutes-long video. In the current frame, a pantry filled with food can be seen. Below the video pane, there is a outline pane (bottom left), which summarizes the video by listing the scenes and errors in the video. On the right top of the figure is a tool pane, which allows users to trim or change the playback speed of the selected video clip. On the right to the tool pane is a search pane with a text field, where users can input the search query of both visual objects or narration. Finally, below the tool pane and the search pane is the audiovisual script, which shows a narration transcript segmented by scenes, scene descriptions, and highlighted visual errors such as blurs and bad lightings.}
\end{teaserfigure}

%%
%% This command processes the author and affiliation and title
%% information and builds the first part of the formatted document.
\maketitle

\section{Introduction}{
People create videos to share their experiences and expertise.
% with a broad audience. 
To create a compelling video, creators first capture raw video footage then edit it to remove irrelevant or low-quality footage and add effects. 
For instance, creators speed up repetitive actions (\textit{e.g.}, walking or cleaning), remove long pauses in their speech, and cut blurry footage due to camera shakes.
With current video editing tools (\textit{e.g.}, Premiere~\cite{premiere}, Final Cut Pro~\cite{finalcut}, Descript~\cite{descript}), creators first need to \textit{visually inspect} the video footage and the corresponding video timeline~\cite{premiere,finalcut} or transcript~\cite{descript}, to identify edit points.
% Requiring visual inspection makes video editing inaccessible for a growing number of blind and low vision (BLV) video creators~\cite{seo2021understanding} who author and share general-purpose and accessibility-focused videos including vlogs, reviews, and tutorials.
Just as visual inspection presents an accessibility barrier for BLV creators authoring static visuals (\textit{e.g.}, photos~\cite{bennett2018teens}, presentations~\cite{peng2022diffscriber}, documents~\cite{das2022co11ab}),
% Similar to accessibility barriers encountered by BLV creators authoring static visual media (\textit{e.g.}, photos~\cite{bennett2018teens}, presentations~\cite{peng2022diffscriber}, documents~\cite{das2022co11ab}), 
relying on visual inspection makes video editing inaccessible to a growing number of blind and low vision (BLV) video creators~\cite{seo2021understanding} who author and share general-purpose and accessibility-focused videos including vlogs, reviews, and tutorials online.
While prior work explored how to make videos accessible to BLV audience members ~\cite{liu2022crossa11y,pavel2020rescribe,liu2021what}, and how to make video editing more efficient for sighted creators~\cite{huber2019bscript,Berthouzoz2012,chi2013democut}, existing work has not yet explored how to make video editing accessible to BLV creators.} 
To better understand BLV creators' current video production strategies and challenges, we interviewed 8 BLV video creators and analyzed 24 videos about screen reader-based video editing.
While BLV creators devised creative techniques to film and edit their videos such as describing the visual content during video capture, and editing the video using audio editing tools, the creators reported that video editing remained challenging due to 
% the following accessibility barriers of videos themselves and video editing tools: 
four core accessibility barriers of videos and video editing tools: lack of access to the visual content of the video (\textit{e.g.}, settings, objects), lack of access to the visual quality of the video (\textit{e.g.}, lighting, blurriness), lack of efficient ways to navigate to different parts of the video, and limited screen reader support (\textit{e.g.}, deeply nested menus or icons listed as ``button'').
% efficiently navigate to different parts of the the video without watching the video through, and (4) 
% % In a formative study, BLV creators reported the following challenges in non-visual editing.
% First, BLV creators lack access to the visual content of the video. 
% Second, BLV creators cannot assess the visual quality of the video. 
% Third, BLV creators lack efficient ways to navigate to different parts of the video. 
% Finally, BLV creators find it difficult to use the existing editing tools with limited screen reader support. 
As a result, BLV creators reported that they either recruited sighted collaborators to review and edit their videos, or uploaded their original video recordings without editing the footage to their preferred level of polish.
% as much as they would like to.
% without editing to their preferred level of polish.
% often upload original recordings without any edits or ask for help from the sighted people to review and edit the videos.

To address accessibility barriers of current video editing tools, we present \sysname{}, a system that supports accessible video editing via \textit{audio-visual scripts} (\autoref{fig:teaser}). ~\sysname{}'s audio-visual script (\autoref{fig:teaser}B) features a transcript of the narration in the video {(\textit{e.g.}, ``First of all,...'')} overlaid with information about the visual content in the video {(\textit{e.g.}, ``Scene 7: A pantry full of food...'')} and visual errors in the video (\textit{e.g.}, ``Camera blur'').
We align the audio-visual script to the video such that BLV creators can directly review, navigate, and edit the video via text.
{As creators play the video, ~\sysname{} surfaces visual information by alerting creators to scene changes and visual errors using audio notifications. \sysname{} also allows creators to inspect objects in the current frame using the ``Inspect'' feature.}
To facilitate efficient navigation, ~\sysname{} features an outline (\autoref{fig:teaser}C) and search feature (\autoref{fig:teaser}E).
~\sysname{}'s outline (\autoref{fig:teaser}C) of key scenes and errors lets creators skim to gain a high-level overview of the video or click to navigate to the corresponding point in the script and video. ~\sysname{}'s search (\autoref{fig:teaser}E) lets creators navigate the video by searching for visual objects (\textit{e.g.}, ``microwave'') or transcript words of interest.

{To assess \sysname{}, we conducted a within-subjects study with 12 BLV editors comparing \sysname{} to the their existing workflows and invited 3 BLV creators to edit their own footage using \sysname{}. In the within-subjects study with 12 BVI editors, creators reported lower mental demand and greater confidence in their final video when editing videos with ~\sysname{} compared to using their own video editing tools (\textit{e.g.}, Reaper, a timeline-based editor, and FFmpeg, a command-line tool). All creators expressed that they wanted to use the tool in the future as it helped them efficiently review their video footage and identify visual errors. BLV creators editing their own footage with \sysname{} used \sysname's visual descriptions to efficiently recall what they filmed, and \sysname's error detection to remove visual errors. After using \sysname{} to edit their own footage, creators reported that ~\sysname{} would enable them to edit more videos without assistance, thus decreasing the time required to produce videos and empowering them to create new types of videos with more diverse content and styles.} 
% Creators that using ~\sysname{} will enable them to edit more videos without assistance, thus decreasing their time and effort to produce videos, and powering them to create videos with diverse content and styles.}

% To assess the feasibility of \sysname{} for enhancing the editing experience for BLV creators, we conducted a within-subjects study with 12 BLV video editors. Creators reported \revised{lower} mental demand and expressed greater confidence in their final results when editing videos with ~\sysname{} compared to their own video editing tools (\textit{e.g.}, Reaper, a timeline-based editor, and FFmpeg, a \revised{command-line tool}). All creators expressed that they wanted to use the tool in the future as it helped them efficiently review their video footage and identify visual errors.

% To further understand how BLV creators would use ~\sysname{} in practice, we invited 3 BLV creators to edit their own footage using \sysname{}. BLV creators used the system to efficiently recall what they filmed and recover from visual errors. \revised{In the interview, creators expressed} that using ~\sysname{} will enable them to edit more videos without assistance, thus decreasing their time and effort to produce videos, and powering them to create videos with diverse content and styles.
% make their editing process more independent
% In particular, 
% \xac{is there any other interesting findings about how they used AVscript?}\amy{+1 especiallya bout the visual content}
Our work contributes:
\begin{itemize}
    \item A formative study revealing current practices and unique challenges of video editing by BLV creators.
    \item Design and development of \sysname{}, a novel system that uses \textit{audio-visual script} to improve accessibility in reviewing, navigating, and editing videos.
    \item Two user studies demonstrating how BLV creators leverage \sysname{} to edit given videos and their own videos.
\end{itemize}

\section{Related Work}
Our work builds upon previous work in accessible authoring tools (Section ~\ref{RW4}), video accessibility (Section ~\ref{RW3}), text-based editing tools for audio and visual media (Section ~\ref{RW1}), and interaction techniques for video navigation (Section ~\ref{RW2}). 
% \amy{I think we should reorder this to be: accessible authoring tools, video accessibility, then text-based editing and video navigation (these final two sections could possibly be combined). I think the accessibility ones outline the background of the problem a bit more, whereas the video ones really relate to specifics of our system design}

\subsection{Accessible Authoring Tools}\label{RW4}
Prior research has explored how creators with visual impairments currently author photos~\cite{adams2016blind, brady2013investigating}, drawings~\cite{kurze1996tdraw}, documents~\cite{villaverde2014facilitating}, websites~\cite{li2019editing}, presentations~\cite{schaadhardt2021understanding, peng2022diffscriber}, audio~\cite{saha2020understanding}, and videos~\cite{seo2021understanding}. Such work identified that authoring tools for visual content remain inaccessible because they do not convey visual information about the content the creator is authoring (\textit{e.g.}, framing of a photo~\cite{adams2016blind}, layout of a website~\cite{li2019editing}).
% that an author with a visual impairment is creating.
Thus, prior work created tactile displays to make authoring websites~\cite{li2019editing}, documents~\cite{avila2018tactile}, and maps~\cite{shi2020molder} accessible. While tactile displays allow creators with visual impairments to access visual content, creators must have access to an embosser or laser cutter to print tactile sheets for the initial and revised visual designs. 
% the adoption of tactile displays is limited due to physical constraints. 
To provide access to visual designs and revisions without tactile displays, Peng et al. generated visual descriptions that let presenters with visual impairments obtain information about the content, layout, and style of their slides~\cite{peng2022diffscriber}.
Prior work also explores how to support collaboration between creators who are sighted and creators with visual impairments (\textit{i.e.} mixed-ability collaboration) by providing descriptions of revisions~\cite{peng2022diffscriber}, and access to awareness information that describes who was authoring what during collaborative editing~\cite{das2022co11ab,lee2022}. 
% More recent work investigated approaches to make collaborative text document authoring accessible by providing awareness information ~\cite{das2022co11ab,lee2022}, and to make visual design authoring (\textit{e.g.,} presentation slides) accessible by generating change descriptions that allow blind presenters to obtain information about the visual content they are authoring (\textit{e.g.,} slide content, layout, and style)~\cite{peng2022diffscriber}. 
While these tools make authoring static visuals including text documents~\cite{das2022co11ab,lee2022} and visual designs~\cite{peng2022diffscriber}, accessible for creators with visual impairments, we explore how to make video authoring accessible to creators with visual impairments by representing the dynamic visual and audio content of a video as text.
% in this work, we expand the exploration of accessible authoring process by turning videos ---media with dynamic visuals--- into accessible interactive descriptive elements.

\subsection{Video Accessibility}\label{RW3}
Creating an accessible tool for authoring videos is challenging partially due to the inaccessibility of videos themselves. 
Videos are inaccessible to BLV audiences when the visual content in the video is not described by the audio (\textit{e.g.,} travel videos with scenic shots set to music)~\cite{liu2022crossa11y,liu2021what,peng2021say}.
To make videos accessible, video creators~\cite{pavel2020rescribe}, volunteers~\cite{youdescribe}, or professional audio describers~\cite{3playmedia} add audio descriptions to describe important visual content that is not understandable from the audio alone. 
% However, creating audio descriptions is a skilled and time-consuming task that is especially challenging for novices.
As authoring audio descriptions is challenging, prior work developed tools that help creators gain feedback on audio descriptions~\cite{saray2011adaptive, natalie2020viscene}, respond to audience requests for descriptions~\cite{youdescribe}, optimize descriptions to fit within time available~\cite{pavel2020rescribe}, and recognize mismatches between audio and visuals to add descriptions as they capture~\cite{peng2021say} or edit~\cite{liu2022crossa11y} videos. Beyond helping creators author accessible videos, prior work makes inaccessible videos accessible on demand by generating automatic~\cite{wang2021toward} or interactive~\cite{peng2021slidecho, huh2022cocomix} visual descriptions. 
While such prior work provides BLV audience members access to visual content in videos, the prior approaches were designed to make videos accessible for consumption rather than authoring, such that they lack important information required for video authoring tasks (\textit{e.g.,} lighting, camera stability). 
We investigate how to improve the accessibility of video authoring by providing text descriptions of both the visual content and quality of videos.
% We investigate how to improve the accessibility of video editing by providing text descriptions of both the visual content and quality of videos that let BLV creators understand their video footage and decide how to edit it. 
% In this \revised{paper}, 
% We investigate improving the accessibility of video editing with text descriptions of the visual content and quality of videos to let BLV creators efficiently understand their video footage and decide how to edit it.
% video content and decide how to edit.
% , but also help decide where and how to edit the materials.

\subsection{Text-based Audio and Video Editing}\label{RW1}
Audio and video editing is time-consuming as it requires multiple iterations of reviewing footage to find edit points, navigating to the edit points, and applying edits~\cite{chi2013democut}. To improve the efficiency of editing audio and video footage, prior work introduced text-based editing, which allows users to edit audio or video as they would a text document by time-aligning the words in the speech transcript with words in the audio~\cite{shin2016, rubin2013content, huber2019bscript, descript, reductvideo, imvidu,truong2016quickcut, Leake2017, Berthouzoz2012, fried2019}. Researchers further improved the efficiency of text-based editing by: highlighting pauses or repeated words in the transcript~\cite{rubin2013content}, suggesting opportunities for B-roll~\cite{huber2019bscript}, and matching voice-over recordings with relevant narrated video segments~\cite{truong2016quickcut}.
In addition, prior research used text-based editing to improve the quality of the video output by creating seamless transitions when cuts or dialogue changes occur in talking head videos~\cite{fried2019}, dialogue-driven scenes~\cite{Leake2017}, and interview videos~\cite{Berthouzoz2012}. 
% While 
However, existing text-based editing tools were designed for sighted video editors who can 
% visually skim the text to navigate efficiently and 
visually inspect video footage to identify editing opportunities, and visually skim the text transcript to navigate efficiently.
% decisions.
% , we explore approaches for making text-based editing accessible including integrating visual content and quality into the script, improving non-visual skimming via an outline, and making the tool screen reader accessible.
% accessible text-based editing tool by integrating visual content and quality into the script, improving non-visual skimming via an outline, and making the tool screen reader accessible.
% , facilitating non-visual skimming and providing access to the visual content and quality has not yet been explored. 
% it has not yet considered how to describe visual content and quality to blind editors, or to help editors non-visually skim the content. 
We explore how to make text-based video editing accessible by 
% Thus, we build upon prior text-based video editing work, but newly explore 
integrating visual content and quality into the speech transcript (i.e. creating an audio-visual script), improving non-visual skimming via an outline, and making editing operations screen reader accessible.
% so that BLV creators can not only use verbal information for editing videos but also locate visual information which is essential in video editing.
% However, as such prior work has not designed tools for blind video editors, it has not considered how to describe visual content and quality to the editors. 

% In particular, Rubin et al. highlighted pauses and repeated words in the transcript to make it easier for sighted authors to visually skim for points to edit~\cite{rubin2013content}. B-Script~\cite{huber2019bscript} allows users to add B-roll clips and change their position and duration based on the transcript.
% To further leverage scripts, pieces of work analyzed the transcript to produce a seamless output when dialogue changes or cuts occur in talking head videos~\cite{fried2019}, dialogue-driven scenes~\cite{Leake2017}, and interview videos~\cite{Berthouzoz2012}. Researchers have also analyzed transcripts to add appropriate visuals to podcast~\cite{xia2020crosscasts}, recommend B-rolls and their positions in a video~\cite{huber2019bscript}, or match voiceover recordings with relevant video segments in narrated videos~\cite{truong2016quickcut}. 

\subsection{Video Navigation Interaction Techniques}\label{RW2}
{Traditional video players such as Premiere~\cite{premiere} and Final Cut Pro~\cite{finalcut} and editors require navigating the video using a timeline. However, timeline-based navigation is challenging as video creators and consumers need to scrub back and forth to find content of interest. 
% \amy{add main approaches used to be more descriptive}
% Navigating videos is one of the challenges in watching videos since users need to scrub back and forth to find the content of interest~\cite{pavel2014digests}. 
To help video consumers skim and navigate to content of interest, prior work introduced approaches to navigate videos based on transcripts~\cite{kim2014data, pavel2015scene, pavel2016vidcrit}, high-level chapters and scenes~\cite{kim2014crowd, chi2012mixt, fraser2020, yang2022catchlive, pavel2014digests, truong2021makeup,pavel2015scene}, or key objects and concepts~\cite{chang2021ruby, liu2018concept, peng2021slidecho}.
% In addition to editing videos using the transcript, prior research also explored navigation using transcripts so that users can search, skim, or browse the video for a particular target efficiently~\cite{kim2014data, pavel2015scene, pavel2016vidcrit}.
While transcripts help users efficiently search for words used in the video~\cite{kim2014data, pavel2015scene, pavel2016vidcrit}, they can be difficult to skim as they are often long, unstructured, and contain disfluencies present in speech~\cite{pavel2014digests}.
To video consumers skim and navigate videos more efficiently, prior work segmented videos into high-level segments (i.e. scenes or chapters) and let consumers browse these segments based on representative thumbnails or text descriptions~\cite{kim2014crowd, chi2012mixt, fraser2020, yang2022catchlive, pavel2014digests, truong2021makeup,pavel2015scene}. 
% is to create chapters or scenes of a video. For example, YouTube introduced a chapter feature where a video is segmented into meaningful sections with titles~\cite{youtubechapter} and is displayed with thumbnails during the search. 
Prior work segmented videos into chapters or scenes by using the transcript to automatically segment the video based on transcript topics~\cite{fraser2020, yang2022catchlive, pavel2014digests, truong2021makeup}, crowdsourcing to annotate segmentation points~\cite{kim2014crowd}, or metadata such as command logs to segment based on interactions~\cite{fraser2020,chi2012mixt,pavel2016vidcrit}. 
As it is particularly challenging for screen-reader users to skim text~\cite{ahmed2012read}, we similarly segment the video to create an outline of important moments (\textit{e.g.,} scene descriptions, visual errors) such that readers can quickly navigate our the audio-visual script using the outline.
% as they would a Google Docs outline. 
We also build on prior work that uses low-level features in the video (\textit{e.g.,} keywords~\cite{chang2021ruby}, presentation slide elements~\cite{peng2021slidecho}) to facilitate search, as we similarly enable search via speech and visual objects to help BLV creators locate footage to edit.}

\section{Formative Study}
Prior work explores practices of content creators with visual impairments creating media such as photos~\cite{adams2016blind,bennett2018teens}, drawings~\cite{kurze1996tdraw}, documents~\cite{villaverde2014facilitating}, and audio~\cite{saha2020understanding}, and explores community aspects of YouTube content creation such as high-level motivations for content creation and engagement with viewers~\cite{seo2021understanding}, but existing work has not yet explored BLV creators' unique practices and challenges in video editing.
% (\textit{e.g.}, high-level motivations for content creation and collaboration)
% documents  and audio~\cite{
% Prior work exploring the community of BLV creators on YouTube 
% One prior work~\cite{seo2021understanding} has explored the community of BLV creators on YouTube, yet it did not cover the e
% To the best of our knowledge, no existing work sought to understand BLV creators' unique practices and challenges in video editing.
To understand how BLV video creators edit videos, we analyzed {24} YouTube videos\footnote{See Appendix C for details on our video collection approach} {(V1-V24)} by 20 BLV creators about their video editing processes and interviewed an additional 8 experienced BLV video creators (P1-P8, Table~\ref{tab:participants}) about their video editing motivations, current practices, and accessibility barriers\footnote{See Supplemental Material for the full list of questions}. 
% \mina{revisit here} 
Participants were recruited using mailing lists and compensated \$ 20 USD for the 1-hour semi-structured interview. We transcribed\footnote{https://clovanote.naver.com} the YouTube videos and interview recordings, then two researchers first independently coded all videos using open-coding~\cite{hartson2012ux}, then met to resolve conflicts and update codes. 
% Then, the two coders went through the codes of each video to resolve any conflict in the code. 
% The list of codes was consolidated after reaching a consensus and updating the codes.
Then, the same two researchers used affinity diagramming~\cite{holtzblatt1997contextual} to group the codes into themes: goals for editing videos, strategies for editing videos, and challenges in editing videos.

\subsection{Findings: Goals for video editing}\label{form_results}
Interview participants reported that their motivation for editing was to make videos more engaging (6 participants), or tailor videos to a specific audience (2 participants). As P2 summarized: \textit{``I only keep the highlights [...] because short and snappy videos are more popular''} (P2).
When editing, 5 participants mentioned that they polished their videos by editing out visual or audio mistakes, and 4 participants highlighted that they make videos concise by removing unimportant footage. 
For example, participants mentioned they removed audio mistakes like `um's and `ah's in the video (P2, P3, P5), pauses in the speech (P5, P7), or answering an audience question incorrectly (P3).
While participants currently edited primarily via the video's audio track, they were often creating videos for a broad audience:  \textit{``My video is not just for people with visual impairments. For sighted viewers, I want to make sure nothing is visually awkward''} (P7).
P2 added that editing the visuals is particularly crucial for BLV video creators as they often make mistakes while capturing video (\textit{e.g.}, filming with lights turned off), and re-filming can be a burden. 
Finally, in addition to cutting out unimportant footage and mistakes, participants wanted to capture viewer attention by adding music and intros to their videos.

\subsection{Findings: Strategies for video editing}\label{practices}

\ipstart{Describing visual content and mistakes} BLV creators are unable to recognize the visual content in a video (\textit{e.g.}, objects, actions, background setting) unless the visual content is understandable from the audio alone (\textit{e.g.}, described by a narrator, or accompanied by sound)~\cite{liu2021what}. 
Thus, most participants mentioned that they verbally described visual content (\textit{e.g.}, where they are and what they are doing) as they filmed their video.
Participants reported a dual benefit to visual descriptions: identifying visual content while editing, and making the final video more accessible to BLV audience members.
% Participants reported that their visual descriptions helped them both identify visual content while editing \amy{(PX, PX, PX)}, and make the final video more accessible to BLV audience members \amy{(PY, PY)}. 
In addition to describing visual content, P3, P4, and P8 added verbal cues for editing when they made mistakes during filming. For example, P8 explained \textit{``When I drop something, I'd say out loud `Don't use the earlier part' so that I will easily remove it later''}.
% or they would say out loud \textit{``three, two, one, cut here''} to indicate where they want to cut later.
P7 dealt with a lack of information about the visual content and quality by focusing on the audio: \textit{``Because I cannot check the visual quality of the footage, I am very picky about the audio. If there is some traffic noise, I don't use that part.''} 
P8 renamed his video files with visual descriptions or editing cues so that he could locate relevant clips without playing the video.

However, creators' reported that their descriptions were inaccurate or incomplete when they were not aware of all relevant mistakes (\textit{e.g.}, bad lighting, blur, poor framing or composition) or visual content.
P2 recalled that once \textit{``When I was pointing at an object describing it, it wasn't there!''}. 
Creators also shared that constantly describing visual content during filming took attention away from being creative (V19) and forced them to replace their audio track with music when they did not want the narration to be included in the final video (P7).

\ipstart{Identifying accessible video editing tools} To edit videos, seven participants used timeline-based editing tools (\textit{e.g.}, Final Cut Pro) and one participant used FFmpeg, a command-line tool (P4). 
% The selection of the editing tools was largely dependent on the screen-reader accessibility of the tool itself and the documents. 
Participants noted that video editing tools were largely inaccessible: \textit{``Finding an accessible editing tool in the first place is difficult. Very few tools are accessible themselves and also have accessible documents or tutorials.''} (P4). Participants identified accessible video editing tools via other BLV creators and then learned how to use these tools with a screen reader via text tutorials, videos aimed at BLV editors, and official documentation. 
Even with the most accessible timeline-based tools, participants reported that the menus, buttons, and sliders were often unclickable with a screen reader or not properly labeled (\textit{e.g.,} only reading `button' instead of describing the function). 
In addition, such tools have complex menus that are difficult to navigate with a screen reader: \textit{``Having no access to GUI, I have to continuously tab to locate the button of my interest. This becomes tedious because video editing is a complex task''} (P1). 
% The final challenge (C4) is that the conventional video editing tools have limited support for screen-reader, limiting the tools and features that BLV creators can use. 
% BLV creators have to spend a lot of time finding an accessible video editor and learning how to use it with a screen reader. P4 mentioned ``Finding an accessible editing tool in the first place is difficult. Very few tools are accessible themselves and also have accessible documents or tutorials.'' Some of the menus, buttons, or sliders of the editing tool were not properly labeled (\textit{e.g.,} only reading `button' instead of describing the function), or unclickable with a screen reader. 
% P1 noted ``Having no access to GUI, I have to continuously tab to locate the button of my interest. This becomes tedious because video editing is a complex task that involves many actions - when writing an email I don’t need to switch cursors multiple times.'' \amy{** how did they learn **, many video editing videos on YouTube but many are not intended for blind learners, YouTube, FFmpeg API, ask community to choose software}

% \cite{saha2020understanding}

\ipstart{Navigating videos linearly} While most BLV creators edited their videos with timeline-based tools, the visual aids that these tools provide for sighted creators to browse, skim and select video footage (\textit{e.g.}, visual thumbnails to preview video content by hovering, audio waveforms to preview start and end of speech) are not accessible to BLV creators.
As a result, all participants reported that they review and edit videos by first watching the entire video all the way through, and either editing as they go (6 participants) or noting timestamps to edit later (P1, P4).
P7 noted that he usually watched the entire clip several times because he cannot jump from place to place in the video.
To avoid re-watching long videos in order to find content of interest, participants commonly filmed multiple short clips: \textit{``Because navigating within a single clip is so tedious, I never create a long clip.''} (P8).

\ipstart{Recruiting sighted collaborators}
BLV creators commonly sought out assistance from sighted people for filming, editing, and reviewing the final video (V1, V3, V6, V19, P2, P3, P7, P8). 
% Three participants used their own verbal cues to locate the part where they made mistakes during filming. 
% P8 noted ``When I drop something, I'd say out loud ``Don't use the earlier part'' so that I will easily remove it later. ''
% Getting assistance from sighted people in certain steps of editing was common (V1, V3, V6, V19, F2, F3, F7, F8). 
For example, the creator of V3 has her sighted husband set up a camera for filming, make video intro templates, and apply color correction. The creator of V1 and V6 recently hired an editor for even basic editing tasks as editing takes too long causing back pain.
% the creator of V3 gets help from her sighted husband with setting the camera position for filming, making intro templates, and applying color corrections, while the creator of V1 and V6 gets help from a hired editor with basic editing tasks because editing takes too long causing back pain.
Before publishing their video, 4 participants (P2, P3, P6, and P8) wanted a sighted person with or without video editing experience to view the video and provide a sense of how an ``average viewer'' would see it. 
% For example, the creator of V3 said that her husband helps her set up the tripod before filming, make intro templates, and apply color corrections. 
% Also, the creator of V1 and V6 noted that he started to get help from an editor even on the basic editing that he can do because editing for a long time causes too much back pain.
% After finishing editing videos, participants often asked sighted people to review their videos. 
For example, P3 often uses Be My Eyes\footnote{https://www.bemyeyes.com/} or Aira\footnote{https://aira.io/} to ask a volunteer to provide feedback on visual quality (\textit{e.g.,} whether she is centered in the frame and well-lit). 
All interview participants mentioned that they wished to edit videos independently, as sighted assistance is not always available or affordable, and they wanted to gain control over the process as creators.
% P2, P6, and P8 wanted to get a sense of the average viewer before uploading. They mentioned that the reviewer doesn't have to be professional in video editing, and can be anyone with vision. 
% While getting assistance from sighted people in certain steps of editing was common, 
% Still, all interview participants responded that they wish to edit videos independently, as the assistance is not always available or costs extra money, and they wanted to gain full control over the process as creators.

% While filming a video, most participants mentioned that they verbally describe the context (e.g., where they are, what they are doing) because they mostly rely on narration when editing (PX, PX, PX) and the description makes the video more accessible to viewers with visual impairments (PX, PY). Similarly, P3, P4, and P8 added some verbal cues (``Three, two, one, cut here'') to indicate where they want to cut out later. 

% using zoom/screen reader
% - how often, how long it takes

\subsection{Reflection: Opportunities for BLV creator support}\label{challenges}
While BLV creators resourcefully crafted strategies to work around inaccessible video editing tools, creators' remaining challenges (C1-C5) point to opportunities for technical or social support:
% By analyzing current practices of BLV creators, we identified \amy{X} key opportunities for technology to support BLV creators: 
\begin{itemize}
    \item[\textbf{C1.}] Recognizing visual content in a video (\textit{e.g.}, setting, actions) 
    \item[\textbf{C2.}] Assessing the visual quality of a video (\textit{e.g.}, lighting)
    \item[\textbf{C3.}] Accessing editing tool menus with a screen reader
    \item[\textbf{C4.}] Non-linear browsing and skimming of videos
    \item[\textbf{C5.}] Performing visual edits (\textit{e.g.}, color correction)
\end{itemize}

Our formative study indicates that BLV creators currently extend time, effort, creative agency, and social resources to overcome these challenges. For example, by narrating the visual content and noting mistakes as they film (\textbf{C1, C2}), losing and regaining editing task focus to navigate menus (\textbf{C3}), spending time watching a video linearly rather than jumping to the point of interest (\textbf{C4}) and recruiting sighted collaborators for inaccessible or overly tedious tasks (\textbf{C1-5}).
Our work explores how to make video editing more accessible by providing creators' access to video visuals (\textbf{C1, C2}) and more efficient by improving the ability of creators to skim and browse for content of interest (\textbf{C4}), while the remaining challenges (\textbf{C3, C5}) indicate rich opportunities for future research and commercial accessibility improvements.
\section{\sysname{} }
% Informed by the formative study, we designed and developed 
\sysname{} (\autoref{fig:teaser}) makes video editing accessible and efficient for BLV creators with \textit{audio-visual scripts} that let creators navigate and edit based on a text script of visual content, visual quality, and speech.
% tool that describes the video's visual content and visual quality in the script for accessibility. 
We first illustrate how BLV creators can use ~\sysname{} to edit videos through an example use scenario. Then, we describe ~\sysname{}'s interface and the computational pipeline that powers it.
% (\autoref{pipeline}).

\subsection{Editing a video with ~\sysname{}}\label{scenario}
Anna, a YouTube content creator with a visual impairment, filmed a cooking tutorial video to upload to her channel. 
Anna wants to improve the conciseness and quality of her tutorial to make it engaging to viewers, so she imports the tutorial video into ~\sysname{} to edit it.
\begin{figure}[t]
\centering
\includegraphics[width=\columnwidth]{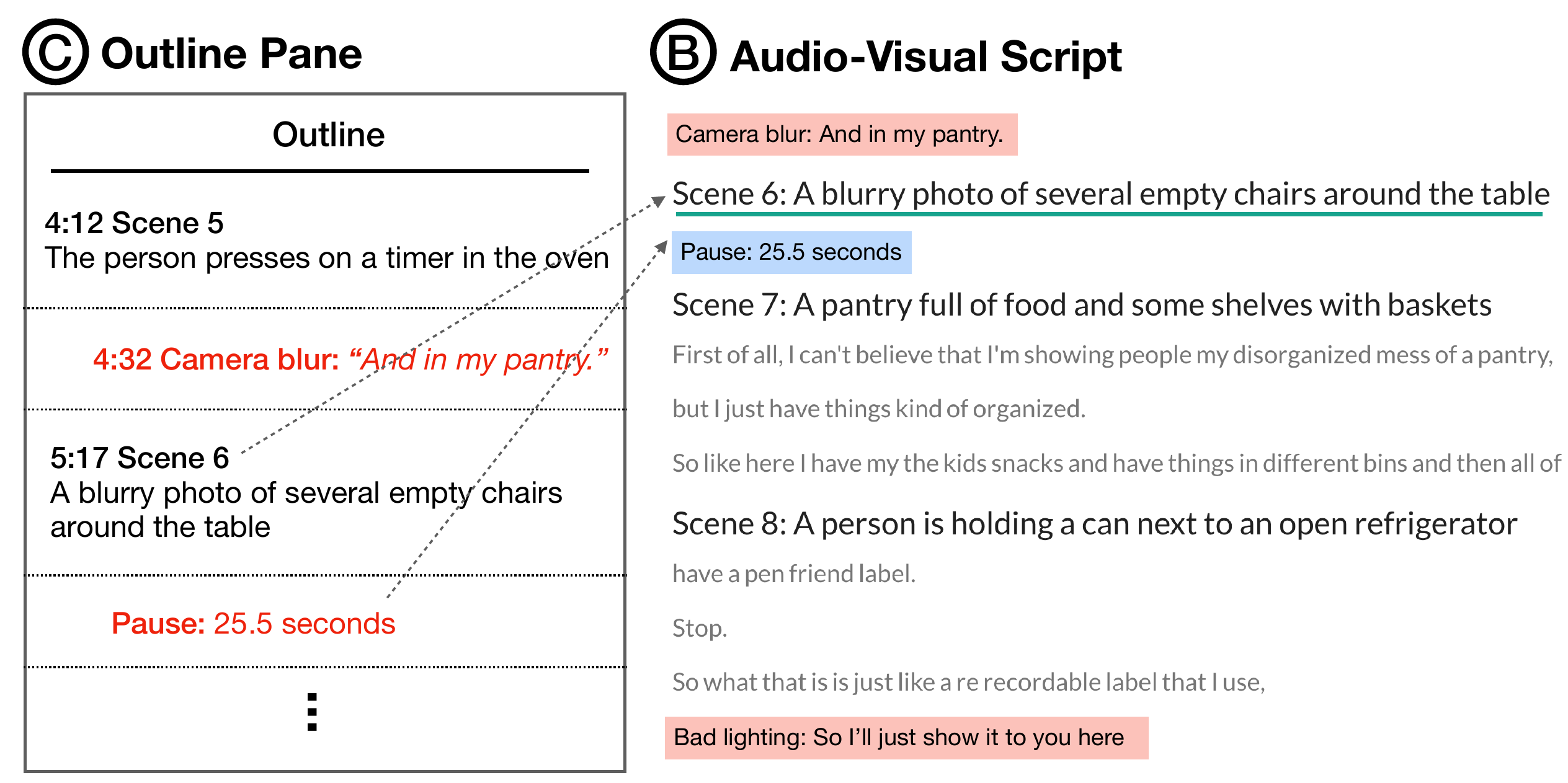}
\caption{~\sysname{}'s outline pane displays a navigable summary of the audio-visual script including the high-level scenes and potential edit points (pauses and visual errors).}\label{fig:outline-pane}
\Description{The outline pane and the audiovisual script are presented in the figure. In both panes, information about scenes 5 to 8 is shown. On the left outline pane, scene descriptions such as ``The person presses on a time in the oven'' and visual errors such as ``Camera blur in 4:32'' are listed. On the right audio-visual script, all of this information is embedded with the narration script.}
\end{figure}
% outline
Using her screen reader, Anna first skims the outline to review her footage (\autoref{fig:outline-pane}). Because the outline summarizes the video with a description of each scene, she quickly recalls the video's content and plans what to edit from the footage. Reading through the outline, Anna notices that the second scene, where she shows her pantry, is over ten minutes long. To shorten the scene, she clicks the item of the outline and jumps to the pantry scene.
% notification, inspect feature
As she plays the video from the start of the pantry scene, Anna hears a notification indicating that there is a visual error (\autoref{fig:video-pane}). To check the error, she pauses the video. As the position of her cursor in the audio-visual script updates alongside the video progress, 
she can easily read the corresponding line in the audio-visual script which indicates there was a camera blur. To learn more about the visuals at that point, she presses the `i' key to inspect the frame. As ~\sysname{} reads out the objects detected in the frame, Anna notices `door' and `hand' and realizes that the camera was shaking as she tried to open the pantry door. To remove the blurry footage, Anna selects the line that contains `camera blur' and deletes the text. 
% search, tool pane
Anna also remembers that she spent a long time silently waiting for the microwave to finish while filming. She searches for `microwave' (\autoref{fig:search}) to find where the microwave appeared in the video and clicks on the relevant result. She shortens the pause by using the `speed change' feature to make the clip two times faster.
% Anna also remembers that there was a long silence when she was using the microwave. To quickly jump to that part of the video, she searches for `microwave' (\autoref{fig:search}). Then, she finds the relevant part from the visual search results and click the result to navigate. To make the pause shorter, Anna uses the `speed change' feature to set the playback speed to two times faster.

\subsection{Interface}\label{interface}

\begin{figure}[t]
\includegraphics[width=0.4\textwidth]{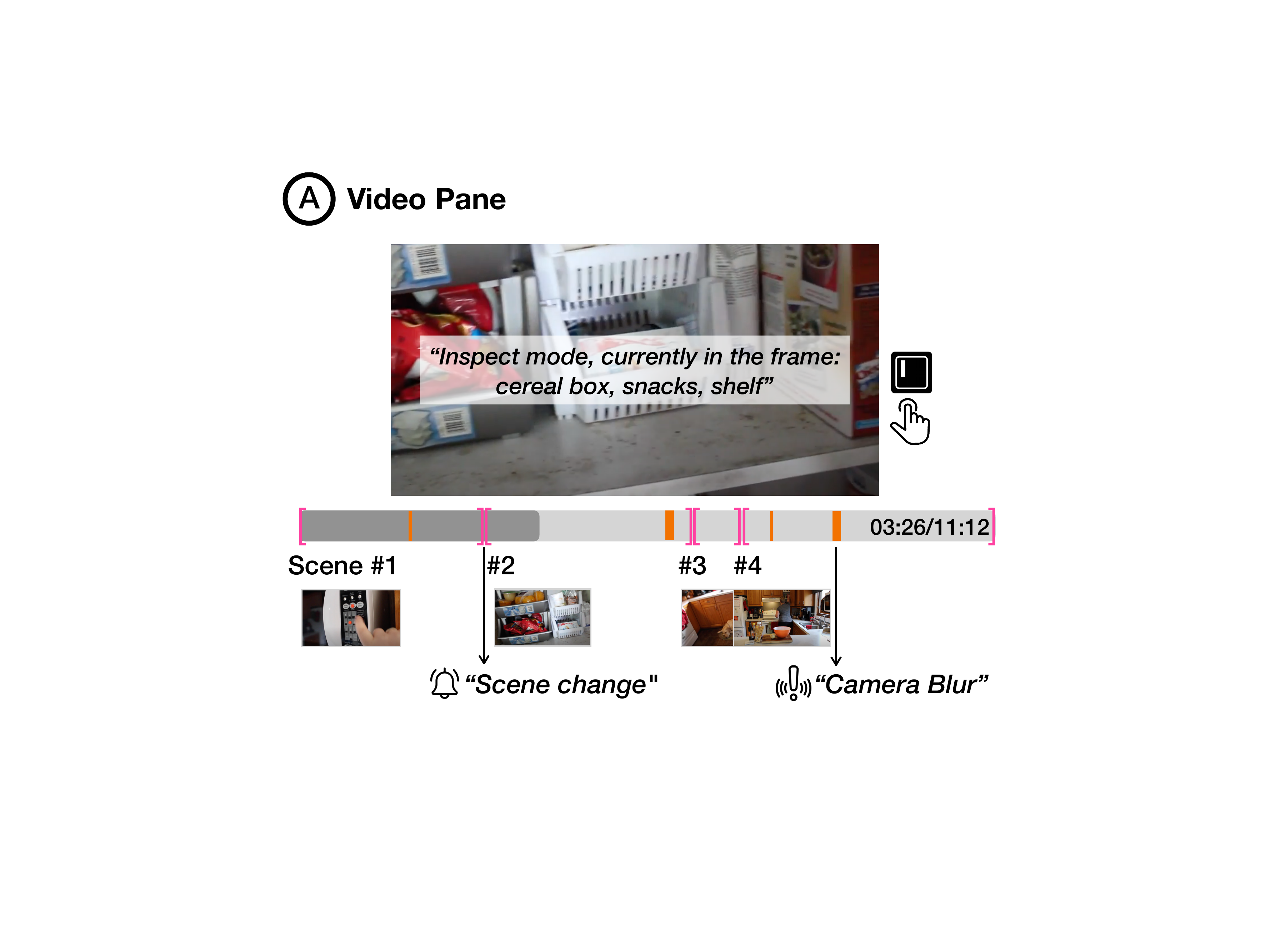}
\caption{\sysname{}'s video pane provides two types of audio notifications: scene change notifications (page-flipping sound) and visual error notifications (warning sound). By pressing the `i' key, users can activate inspect mode to access detected objects in the current frame.}\label{fig:video-pane}
\Description{The close-up of the video pane. Over the video player timeline, notification icons are shown to indicate that this information is provided as users play the video. When the user clicks the `i' key, the following speech is provided: ``Inspect mode, currently in the frame: cereal box, snacks, shelf''}
\end{figure}

The ~\sysname{} interface consists of: a ~\textit{video pane} (\autoref{fig:video-pane}A), a ~\textit{audio-visual script} (\autoref{fig:outline-pane}D), an ~\textit{outline pane} (\autoref{fig:outline-pane}C), a ~\textit{tool pane} (\autoref{fig:search}D), and a ~\textit{search pane} (\autoref{fig:search}E). 

\subsubsection{Video Pane}
The ~\textit{video pane} displays the video and the timeline (\autoref{fig:video-pane}). As the user listens to the video, the system provides sound notifications for the key visual events (e.g., ``Scene Change'', ``Camera Blur''). Users can access visual information in the current frame by pausing the video and pressing the `I' key to activate \textit{inspect mode}, which reads out a list of detected objects in the frame.
% (\textit{Inspect mode}). ~\sysname{} then reads out the list of objects detected in the frame.

\subsubsection{Audio-Visual Script \& Outline Pane}
The ~\textit{audio-visual script}'s \textit{audio-visual script} (\autoref{fig:outline-pane}D) displays the narration and pauses in the video speech along with high-level visual scenes and visual errors. The audio-visual script is aligned to the video, so navigating within the script will navigate within the video (and vice versa), and edits to the script (\textit{e.g.}, selecting and deleting a sentence) are reflected in the video.
The audio-visual script first includes lines that represent each sentence and comma-separated phrases greater than three words in the transcript (\textit{e.g.}, ``First of all...'').
To inform users about the scene changes in the video, ~\sysname{} provides high-level scene headings in the script that summarize the key visual content in the scene (\textit{e.g.}, ``A person is holding a can next to an empty refrigerator''). 
% The heading text summarizes the key visual content of each scene. 
In addition to the scene headings, ~\sysname{} also provides recommendations for potential edit points alongside the text that occurs at that time. For visual errors (highlighted in orange), the system describes the type of error (e.g., ``Bad lighting''), and for long silences (highlighted in blue), the system provides the duration of silence (\textit{e.g.} ``25.5 seconds'').
~\sysname{}'s audio-visual script is designed to enable screen reader users to easily navigate the video at different levels of granularity (high-level visual scenes, narration or pause lines, and words) using key commands (\textit{e.g.}, ctrl/cmd + ~$\rightarrow$/$\leftarrow$ to jump forward or backward by a line).
% The script lines represent each sentence and comma-seperated phrase greater than three words in the transcript.
% : navigating by high-level visual scenes, navigating by line, and navigating by word.
% : users can navigate by high-level visual scenes, 
% In the script, each sentence and comma-separated phrase (> 3 words) is displayed on a different line to enable screen-reader users more easily navigate sentence units using key commands (e.g., ctrl/cmd + ~$\rightarrow$/$\leftarrow$). 
% The cursor in the script is synchronized to the video so that users can easily track the current progress of the video. 
% With script-based editing, ~\sysname{} enables creators to navigate and edit text instead of having to jump back and forth in the video for a fixed amount of duration (D4). 
% \xac{given the similarity to existing features like YouTube's interactive scripts, we should point out what is unique in our design}

In the ~\textit{outline pane} (\autoref{fig:outline-pane}C), the scene headings and recommendations for edit points are listed and sorted in timeline order to provide an overview of the major visual events. By clicking an item in the outline, creators can directly jump to the corresponding part of the video, with an updated cursor position in the script. 

\begin{figure}[t]
\includegraphics[width=\columnwidth]{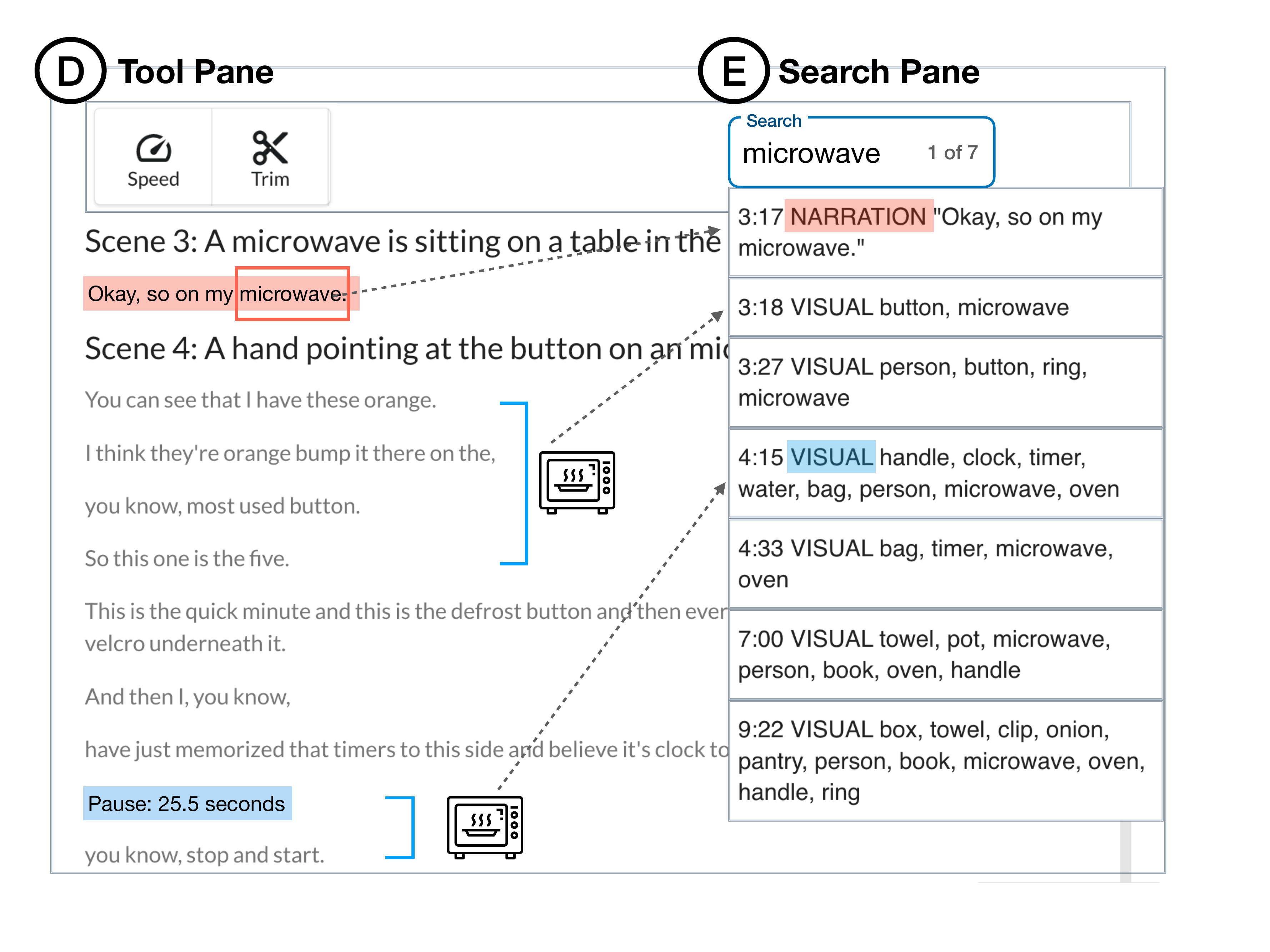}
\caption{~\sysname{} supports search over the transcribed speech and visual objects in the video. BLV creators can skim the results and click on a search result to jump to the corresponding point in the video.}\label{fig:search}
\Description{The tool pane and the search pane are presented in the figure. The user is searching for the keyword ``microwave'' and seven search results are listed, where one of the results is narration search and the other seven results are visual search.}
\end{figure}

\subsubsection{Tool Pane \& Search Pane}
% ~\sysname{} supports three main edit operations: delete, trim, and adjust playback speed. 
To edit the video with ~\sysname{}, the creator selects a segment in the audio-visual script and either presses delete (i.e. backspace) to remove that part of the video, shortens the segment by adjusting the start and end time with the ``Trim'' tool, or changes the playback speed of the segment by using the ``Speed'' tool.
% Instead of removing the entire selection, users can also adjust the start and end time of the selected parts by using ~\textit{trim}, or change the ~\textit{plaback speed}.
When creators have a specific editing target in mind (\textit{e.g.}, a microwave), they can use ~\textit{search pane} (\autoref{fig:search}) to query a speech word, a visual object, or a visual error (\textit{e.g.}, ``microwave'', ``Camera Moving'', or ``Pause''). Then, creators can review and select a search result to jump to the start time of the result in the video and audio-visual script.
% The search box enables creators to search a specific scene from the video with a keyword, and the system lists all scenes where the keyword was mentioned in the narration, or the object of the keyword appeared in the frame. 
% This non-linear way of navigation can be useful to creators who have specific editing targets in mind, or who have limited time to review the entire video before editing.
% \subsubsection{Implementation}label{implementation}
% We implemented ~\sysname{} using React.js, HTML, and CSS for the front-end web interface, and Firebase for the backend server. To r
% Remotion, save the script data and reflect the edits when export...
\subsubsection{{Implementation}}
{
We implemented \sysname{} using React.js, HTML and CSS for the front-end web interface and Firebase for the back-end interface. For embedding a video player, we used Remotion~\cite{remotion} for efficient server-side rendering and parametrization. For \textit{audio-visual scripts}, we used Draft.js~\cite{draft}, a text editor framework for React. We followed the guidelines of W3C~\cite{wai} and tested the compatibility of the \sysname{} with all three major screen readers: \textit{NVDA}, \textit{JAWS}, and \textit{VoiceOver}.
}

\subsection{Computational Pipeline}\label{pipeline}

\begin{figure*}[t]
\includegraphics[width=\textwidth]{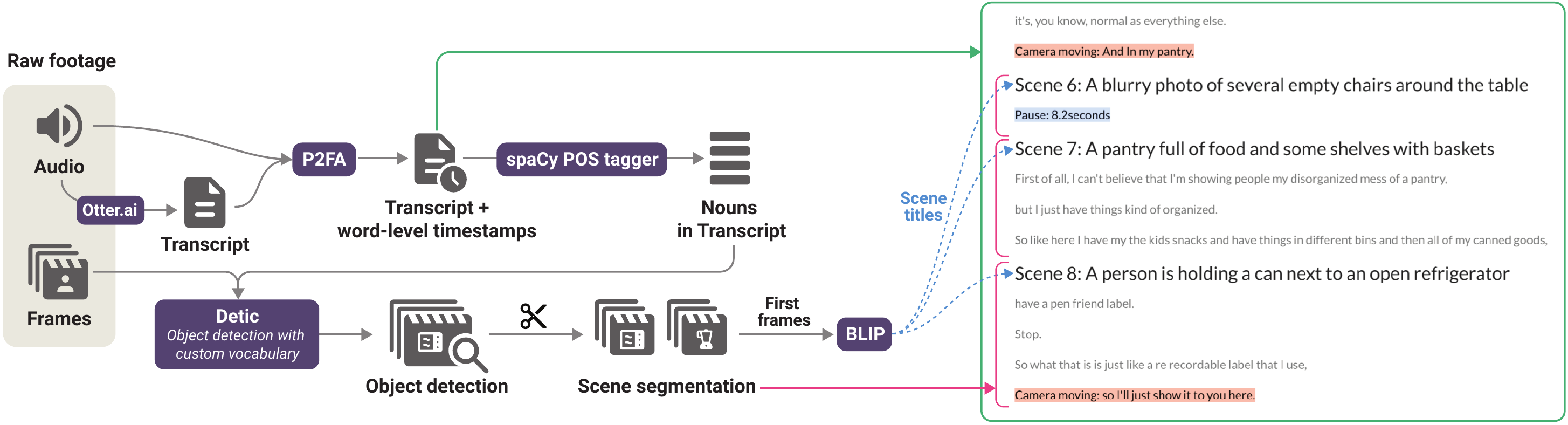}
\caption{Computational Pipeline of creating ~\sysname{}'s audio-visual script from raw footage. It takes two inputs: audio and frames. To generate an aligned transcript, we obtain the transcript from audio using Otter.ai and align using P2FA. To segment the footage into multiple scenes, we first detect objects in each frame with Detic, using the nouns extracted from the transcript as custom vocabulary. Then, we segment the footage when there is a salient change in the objects detected in nearby frames. For each scene, we caption the first frame using BLIP, then use the caption as the scene's title in the audio-visual script. }\label{fig:pipeline}
\Description{The figure shows how audio and visual frames of the video are processed to segment scenes, generate scene descriptions, and detect pauses and visual errors. On the right, the resulting audio-visual script is shown.}
\end{figure*}

% \autoref{fig:pipeline} displays 
~\sysname{}'s computational pipeline (\autoref{fig:pipeline}) transcribes and aligns video speech (Section \ref{transcription}), detects objects and segments scenes (Section \ref{segmentation}), and detects visual errors (Section \ref{recommendation}).
% \revised{All videos were pre-processed before the deployment for the user studies. We downloaded the video and audio using the youtube-dl library~\cite{youtube-dl} and extracted frames using FFmpeg~\cite{tomar2006converting}. Then we passed the data to the pipeline for transcription and audio alignment (Section \ref{transcription}), object detection and scene segmentation (Section \ref{segmentation}), and detection of visual errors (Section \ref{recommendation}).}
% the computational pipeline of~\sysname{}. 
% Given the raw video footage, the system first transcribes and aligns the speech of the video, then segments the video into scenes, and finally detects potential visual errors.
% (See \autoref{fig:pipeline}, left).

\subsubsection{Transcribing and Aligning Speech}\label{transcription}
% edit by word
To enable word-level editing, ~\sysname{} transcribes the video speech using Otter.ai~\footnote{https://otter.ai/home}, then uses P2FA to align each word in the transcript to the corresponding word in the speech.
% to the corresponding speech audio.
Following Rubin ~\textit{et al.}~\cite{rubin2013content}, we use CMU Sphinx Knowledge Base Tool~\cite{sphinx} to obtain word phonemes of out-of-vocabulary words (\textit{e.g.}, the coffee machine name ``Keurig'').
To enable phrase-level navigation and editing, \sysname{} then splits the transcript into sentences and comma-separated phrases that are three words or longer. ~\sysname{} also creates pause segments for any pause longer than three seconds. 
% short segments based on: a word count of greater than 3 , punctuation in the script (commas, periods, exclamation marks, question marks) and a gap between words that is 3 seconds or longer. For each gap longer than 3 seconds, we add a pause segment to the script. 
As widely used screen readers (\textit{e.g.}, VoiceOver, NVDA, JAWS) read the text in HTML `input' elements line-by-line, we place each phrase and pause on a different line for ease of screen reader navigation.
% easier navigation using screen reader keys.

\subsubsection{Segmenting and Labeling Scenes}\label{segmentation}
Using OpenCV we extract frames from the video at a rate of one frame per second.
For each frame, we detect objects in the frame using Detic~\cite{zhou2022detecting} to retrieve visual information of the content for frame inspection, visual search, and detection of major visual changes for scene segmentation.
In our pilot experiments, using all objects detected in the frame resulted in too much irrelevant information passed to the pipeline or presented to the creator (\textit{e.g.,} listing all the objects in the background such as a coffee mug, spoons, forks).
To limit our object detection to objects that are likely to be important, we only detect the objects referred to in the narration. To create a custom vocabulary set, we use Spacy's part-of-speech tagger ~\cite{Honnibal_spaCy_Industrial-strength_Natural_2020} to extract all noun phrases in the transcript (`NN': noun, singular or mass, `NNP': noun, proper singular, `NNPS': noun, proper plural, `NNS': noun, plural). Then we pass the custom vocabulary to Detic~\cite{zhou2022detecting} to detect all instances of each noun in each frame. Detic provides the bounding box for each noun in each frame and a confidence value. We include all objects with a confidence value greater than 0.3.
In the inspect and search mode, ~\sysname{} reads objects in order of the size of their bounding box (largest first).
We segment videos into higher-level scenes by using a sliding window of width 4 to compare \% of similar objects in the 2 frames before and 2 frames after a potential boundary (similar to Haq et al.~\cite{haq2019movie}).
% \amy{make sure this is readable}. 
If a scene boundary occurs in the middle of a phrase boundary, we adjust the scene boundary to match the phrase boundary. We cut short scenes that did not encompass any entire phrase. 
Then, to obtain a description for each scene, we generate the caption of the first non-blurry frame of each scene using a BLIP~\cite{li2022blip}'s pre-trained model (CapFilt-L) with nucleus sampling. {While BLIP produces state-of-the-art image captioning performance, BLIP occasionally misidentified objects, misgendered people, and cited incorrect emotions in pilot experiments.}
% In the later sections, we discuss how the users trusted and noticed the errors in the scene headings.} 

{We evaluated our scene segmentation on two videos (V1 and V2 in Section \ref{sec:eval:materials}) by comparing our predicted scene boundaries to scene boundaries independently labeled by two researchers (Coders A and B who are authors of this paper). We measured percent similarity (i.e. Jaccard Index) between each set of scene boundary labels by dividing the number of matching labels by the total number of labels, considering any labels less than 3 seconds apart as the same.
% For V1, Coder A labeled 22 boundaries and Coder B labeled 25 boundaries, with 12 matching boundary labels (Jaccard Index=34\%). Our segmentation algorithm labeled 15 boundary labels, sharing 10 matching boundaries with Coder A (Jaccard Index=37\%) and 11 matching boundaries with coder B (Jaccard Index=38\%).
% For V2, Coder A labeled 15 boundaries and Coder B labeled 18 boundaries, with 12 matching boundaries (Jaccard Index=57\%). Our segmentation labeled 13 boundaries, with 11 matching Coder A's labels(Jaccard Index=64\%) and 10 matching Coder B's labels (Jaccard Index=48\%).} 
For V1, coders A and B shared 34\% matching boundaries with each other, while our segmentation algorithm shared 37\% and 38\% matching boundaries with coders A and B respectively. 
For V2, coders A and B shared 57\% matching boundaries with each other, while our segmentation algorithm shared 64\% and 48\% matching boundaries with coders A and B respectively. Overall, our segmentation algorithm achieved similar agreement with human coders as they did with each other. When disagreements occurred they typically represented high-quality segmentation boundaries provided at different levels of granularity (\textit{e.g.}, a single segment for adding ingredients vs. three segments for adding flour, water and salt). }

\subsubsection{Detecting Visual Errors}\label{recommendation}
The common components of photo quality that BLV people find difficult to achieve are blur, lighting, framing, and composition~\cite{adams2016blind, brady2013investigating}. 
Among these, ~\sysname{} supports identifying blur and poor lighting, and also considers camera motion blur to support video rather than photo content.
To detect dark lighting, for each frame in the video we reduced the size of the frame to 100x100 to reduce the computation, then classify the frame as ``dark'' if the mean pixel luminescence value falls below an empirically determined threshold of 0.25.
To detect blurry frames, we use the modified Laplacian method ~\cite{pech2000diatom}. For each frame, we first convert the image to grayscale using OpenCV and then compute the variance of Laplacian to calculate the focus score. Then, we classify the frame as ``blurry'' if the focus score falls below an empirically determined threshold of 5.
Using the detection results of each frame, we mark a segment as `dark' or `blurry' when more than three consecutive frames are identified as such. 
Finally, to avoid naively identifying all the camera moving parts (\textit{e.g.,} facing the camera to a different object) as `blurry', we also used the object detection results to detect `camera moving' between scenes (frequent change in the object set over time). For segments that were classified as both ``blurry'' and ``camera moving'', we label them as ``camera moving'' to indicate that the motion blur may make objects in the frame difficult to see. 

{We evaluated our error detection pipeline on two videos (V1 and V2 in Section \ref{sec:eval:materials}) by first creating a set of ground truth labels of visual errors based on existing video editing guidelines~\cite{vlog, Colostate, 101, TTU, handbook, ncsl}. Two researchers first met to group established guidelines into common themes (See Supplemental Material for aggregated guidelines), then researchers seperately annotated edit points for a single video (V1) and met to resolve conflicts and revise the guidelines. 
% To aggregate existing guidelines, Two researchers first grouped common themes from six estab beased on established video editing guidelines~\cite{}.  (see Supplemental Material).
% Two researchers first grouped common themes from e
% , we created a set of ground truth edit points for two purposes: 1) evaluation of the error detection pipeline and 2) evaluation of the quality of videos edited by participants~\ref{output-eval}. To guide us in annotating the edit points, we created a task guideline (see supplementary material) leveraging six different established guidelines for video production. 
% Two researchers first grouped common themes and strategies to create an initial guideline. Then, following the guideline, each of them annotated edit points for a single video (V1) and discussed to resolve conflicts. 
One of the researchers annotated the other video (V2) following the revised guidelines. In total, the ground truth labels included 18 errors for V1 and 15 errors for V2. }
% guideline and the other researcher reviewed and discussed to create final annotations.}
{When compared with the ground truth edit points, \sysname{}'s pipeline achieved high precision and low recall for visual error detection (precision=100\%, recall=38.89\% for V1, precision=87.50\%, recall=46.67\% for V2). The most common error type not detected by our pipeline was a partial blur due to the main object being out of focus as our pipeline only calculates the focus score of the entire frame. One of the reasons for the high precision and low recall is that we empirically set the threshold of \sysname{}'s pipeline low to avoid false notification of errors, or presenting users with too many error suggestions.
}

% \subsection{Design Rationales}\label{rationales}
% % design goals
% Based on the challenges learned from the formative study, we identified four design rationales to improve non-visual video editing experiences. Two of the design goals address the accessibility challenges in non-visual editing (D1, D2), and the other two design goals address how the editing process can be more efficient. (D3, D4)
% can interactively control the reading (D3, D4).
% \begin{itemize}
% \item[D1.] Provide access to the visual content (C1).
% % chapters as headings, inspect mode
% \item[D2.] Inform creators about visual errors (C2).
% % recommendation
% \item[D3.] Support non-linear access to different parts of the video (C3)
% % search feature, outline feature
% \item[D4.] Support content-based editing to let creators work at a higher level (C4).
% % text-based tool
% \end{itemize}
% \xac{what are the 'C's?}

\section{Comparison Study}\label{comparison-study}
We conducted a within-subjects study to examine how ~\sysname{} impacts experienced BLV creators' video editing practice compared to their personal video editing tools. 
% In a 2-hour study, 12 BLV creators with video editing experiences compared ~\sysname{} against their personal video editing tools.

% experienced BLV creators edited videos using both ~\sysname{} and their personal video editing tools.   
% to learn how ~\sysname{} impacts BLV creators' video editing practice.
% To assess how ~\sysname{} compares to BLV creators' existing video editing practice, we conducted a controlled user study to answer:

% \amy{Can just have RQ1 and RQ2 without (1) and (2). I wonder if we can simplify these research questions or if we need them or we can state what we're hoping to learn}
% We explored two research questions:
% \begin{enumerate}
%     \item RQ1. How does ~\sysname{} impact task performance (e.g., efficiency, output video quality, edit accuracy) compared to users’ typical editing practice?
%     \item RQ2. How does ~\sysname{} impact users' perception of the task (e.g., preference, confidence) compared to users’ typical editing practice?
% \end{enumerate}

\begin{figure*}[t]
  \centering
  \includegraphics[width=\textwidth]{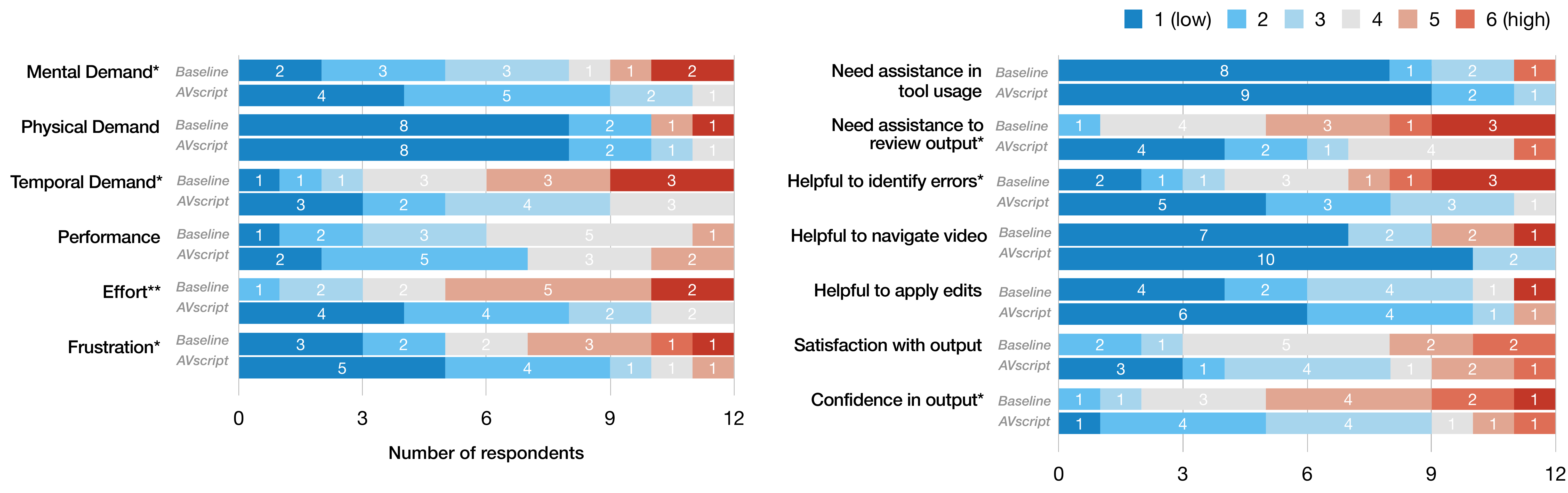}
  \caption{Distribution of the rating scores for the participants' personal editing tools and ~\sysname{} (1 = low, 7 = high). Note that a lower value indicates positive feedback and vice versa. 
  {The asterisks indicate the statistical significance as a result of Wilcoxon text} (~\textit{p} < 0.05 is marked with * and ~\textit{p} < 0.01 is marked with **). \sysname{} significantly outperformed users' own tools in mental demand, Temporal demand, effort, frustration, confidence in the output, independence in reviewing output, and helpfulness in identifying errors.}
  \label{fig:NASA-TLX}
  \Description{The two stacked bar charts display the distribution of the rating scores for the participants' participants' personal editing tools and AVscript. Blue colors are used to indicate lower responses (positive) and red colors are used to indicate higher responses (negative). }
\end{figure*}

\subsection{Method}\label{comparison-methods}

\subsubsection{Participants}
We recruited 12 participants who all had a visual impairment, used a screen reader to access their device, and had prior experience editing videos (8 males and 4 females, aged 28--58) using mailing lists and social media (Table~\ref{tab:participants}). Three of the participants also participated in the formative study (P1, P4, P8).
Among the participants, ten participants had a YouTube channel where they posted videos to the public, while two participants shared their videos privately (\textit{e.g.}, to the company they work for or family and friends). All 12 participants mentioned that they create videos for both sighted and BLV audience members. Participants authored a variety of videos including vlogs, tutorials, product reviews, presentations, and more (see Table~\ref{tab:participants} for a complete list).
To edit videos, 11 participants used timeline-based editing tools (Reaper, Windows Movie Maker, Microsoft Photos, Final Cut Pro, and VideoReDo), and 1 participant used scripting tools (FFmpeg, Python). Reaper is primarily an audio production tool, but it also supports videos. {All participants used one or more of the three popular screen readers (NVDA, JAWS, VoiceOver) which are all compatible with ~\sysname{}}.

\subsubsection{Materials}\label{sec:eval:materials}
We selected three videos from YouTube authored by BLV creators that contained (Table~\ref{tab:videos}): primarily raw video footage with few edits, real-world camera footage rather than screen recordings, and narration in English by the video author.  We selected a short video (V0) for the tutorial. Videos used in the main sessions (V1-V2) were created by the same YouTube creator\footnote{https://www.youtube.com/c/BlindMovingOn} and were selected to be similar in terms of length, amount of narration, and shot changes.
% (X mins, Y mins), the amount of narration (X words, Y words), and shot changes (X shots, Y shots detected using XX). 
For both videos, we only used around the first 11 minutes of the video such that participants could edit the video within the study time. For each video, we did not manually correct algorithmic results except for replacing the incorrect gender identification of the speaker. 

\subsubsection{Procedure}
We conducted a 120-minute remote study on Zoom {where all participants had a 1:1 session with one of the researchers}. 
We first asked participants demographic and background questions about their prior video editing experience. We then gave a 20-minute tutorial on the ~\sysname{} interface in which participants edited V0 to learn system features. Participants then edited one video (V1 or V2) with ~\sysname{} and the other video (V1 or V2) using their existing editing tools (within subjects). The order of system type (their own editing tools vs. ~\sysname{}) and video clips (V1 or V2), was counterbalanced and randomly assigned to participants. {During the task, we answered participant questions about \sysname's screen reader controls and the amount of task time remaining but did not provide any help with understanding or editing the video.} We encouraged participants to take a short break between two sessions. For each interface, we conducted a post-stimulus survey that included three types of questions: NASA-TLX ratings, ratings about the final video output, and ratings about the perceived helpfulness of system operations. {As we did not provide assistance with video understanding or editing during the study, ratings related to \textit{assistance} (Figure \ref{fig:NASA-TLX}) intend to capture participants' perceptions of their ability to use each tool independently.} All ratings were on a 7-point Likert scale.
After the session using ~\sysname{}, the edited video was saved to our server. We also asked the participants to share the output video edited using their personal video editing tools. At the end of the study, we conducted a semi-structured interview to understand participants' strategies using ~\sysname{} and the pros and cons of both ~\sysname{} and their own tools. We compensated participants with a 40 USD Amazon Gift Card. This study was approved by our institution's Institutional Review Board (IRB). 

% \subsubsection{Output Video Quality Evaluation}\label{output-eval}
% \revised{ In addition to collecting the subjective ratings from the participants, }

\subsubsection{Analysis}
We collected the video recordings, the interaction logs, the output videos, and the survey responses to perform both quantitative and qualitative analyses. 
\sysname{}'s interaction logs were collected using Google Firebase~\cite{firebase}.
We reviewed both the session recording and interaction logs to extract the operations participants performed using their baseline video editing tools and \sysname. We triangulated the logs with the output videos to validate the extraction (\textit{e.g.,} comparing the edit points in the video to the edit operations). 
% We did not count the number of primitive actions such as moving cursors without playing a video or subsequent playing/pausing of the same video segment. 
We transcribed the exit interviews and participants' spontaneous comments during the tasks and grouped the transcript according to (1) strategies of using \sysname{} and (2) perceived benefits and limitations of our system.

\subsubsection{Study Limitations} In this study, participants used \sysname{} for the first time and compared it to their own editing tools, which they are already familiar with. Thus, this study neither reveals how long-term use might impact the editing experience of users, nor how participants who have never edited videos before might use the system. We selected the video length (around 11 min) and the editing time provided (around 30 min) to balance providing a realistic use scenario while keeping the study time short, especially as editing is cognitively demanding. 
As a result, not all participants were able to complete editing within the time provided. 

\begin{figure*}[t]
\centering
\includegraphics[width=\textwidth]{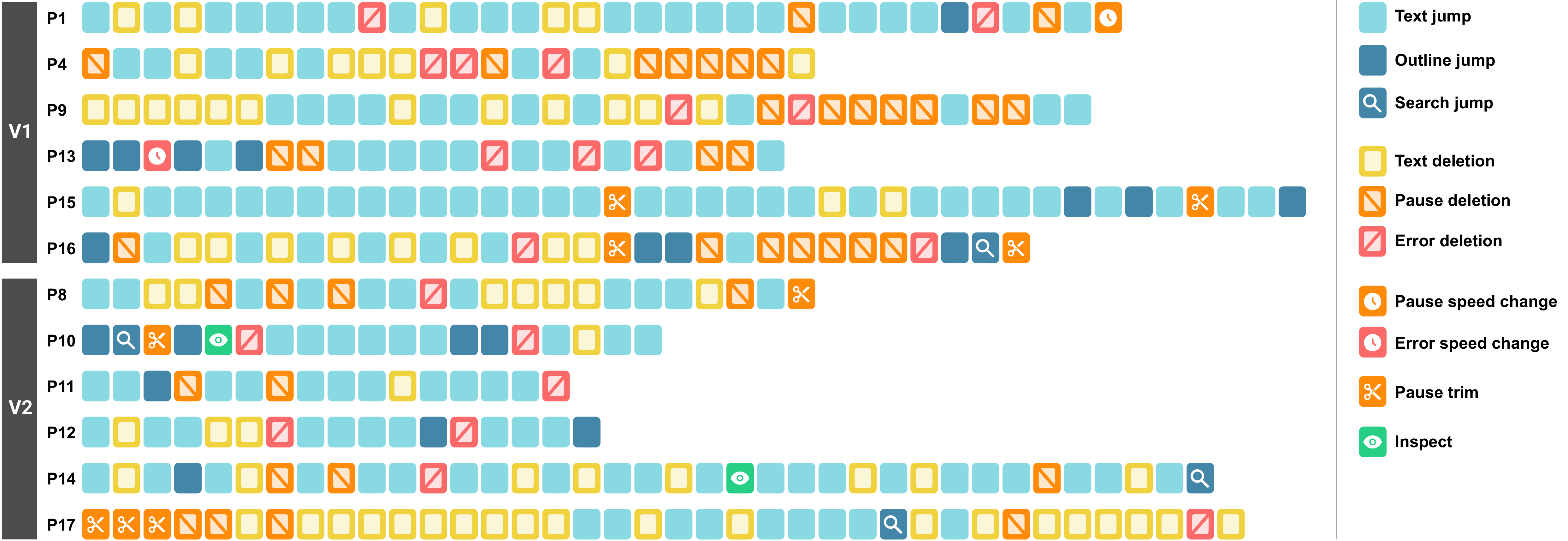}
\caption{Sequences of operations that are relevant to ~\sysname{}'s navigation and editing features by participant (comparison study). Participants' data is grouped by video ID and then sorted by participant ID. Note that trivial cursor movements in the transcript without triggering the video control were not included for brevity.}\label{fig:sequence}
  \Description{The figure shows the pattern of users' interaction when using AVscript including navigations, search, and edits. There is no common pattern among multiple users. }
\end{figure*}

\begin{table*}[t]
\aptLtoXcmd{}{
\small\sffamily
			\def\arraystretch{1.2}
		    \setlength{\tabcolsep}{0.2em}
}
		    \centering
\caption{Summary of the number of edits made by participants using their personal video editing tools (Baseline) and ~\sysname{}. Edit time refers to the time spent on editing each video, and length refers to the duration of the output video. We also report the total number of edits (d: deletion, s: speed change, t: trim, a: audio effect, i: insertion). Note that `trim' has the same effect as deletion but is counted as separate in ~\sysname{} session to distinguish text-based deletion and timestamp-based trim. The video timeline visualizes the position of edits in the video.}
\label{tab:video_stat}
\begin{tabular}{|l|ll|ll|ll|ll|}
\hline
 \multirow{2}{*}{\textbf{PID}} & \multicolumn{2}{l|}{\textbf{Edit Time}} & \multicolumn{2}{l|}{\textbf{Length (mm:ss)}} & \multicolumn{2}{l|}{\textbf{Total \# of edits}} & \multicolumn{2}{l|}{\textbf{Video Timeline}} \\ \cline{2-9} 
 & \textit{Baseline} & \textit{\sysname{}} & \textit{Baseline} & \textit{\sysname{}} & \textit{Baseline} & \textit{\sysname{}} & \textit{Baseline}\hspace{70pt} & \textit{\sysname{}} \\ \hline
& & & & & & & \cellcolor[HTML]{525252}{\textcolor{white}{V2}} & \cellcolor[HTML]{525252}{\textcolor{white}{V1}} \\
1 & 30m & 30m & 8m 2s & 9m 35s & 1 (1d) & 10 (9d, 1s) & \multicolumn{2}{|c|}{\begin{minipage}{.45\textwidth}
      \includegraphics[width=\linewidth, height=3.5mm]{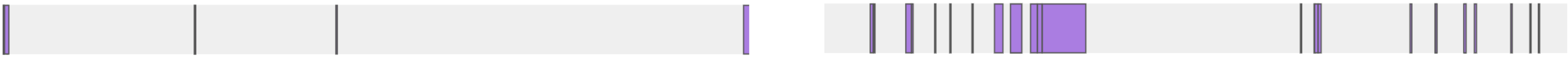}
    \end{minipage}  }  \\
4 & 30m & 30m & 6m 39s & 5m 57s & 2 (2d) & 12 (12d) & \multicolumn{2}{|c|}{\begin{minipage}{.45\textwidth}
      \includegraphics[width=\linewidth, height=3.5mm]{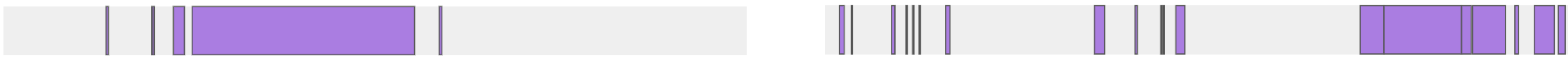}
    \end{minipage}  }  \\ 
9 & 30m & 25m & 6m 58s & 8m 16s & 5 (5d) & 22 (22d)& \multicolumn{2}{|c|}{\begin{minipage}{.45\textwidth}
      \includegraphics[width=\linewidth, height=3.5mm]{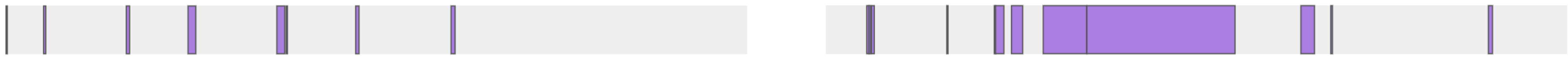}
    \end{minipage}  }    \\
13 & 30m & 20m & 9m 10s & 9m 9s & 4 (3d, 1a) & 8 (7d, 1s) & \multicolumn{2}{|c|}{\begin{minipage}{.45\textwidth}
      \includegraphics[width=\linewidth, height=3.5mm]{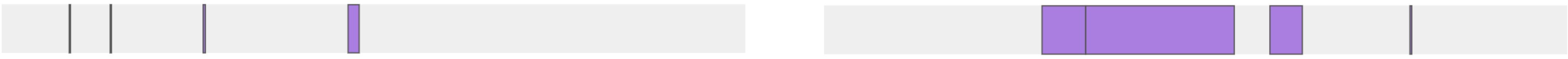}
    \end{minipage}  }    \\
15 & 30m & 30m & 10m 47s & 9m 37s & 6 (6d) & 5 (5d) & \multicolumn{2}{|c|}{\begin{minipage}{.45\textwidth}
      \includegraphics[width=\linewidth, height=3.5mm]{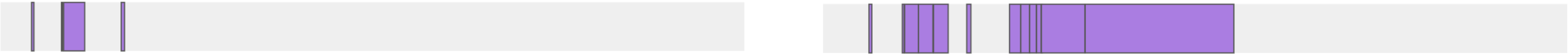}
    \end{minipage}  }    \\
16 & 30m & 30m & 7m 30s & 8m 39s & 5 (5d) & 19 (17d, 2t) & \multicolumn{2}{|c|}{\begin{minipage}{.45\textwidth}
      \includegraphics[width=\linewidth, height=3.5mm]{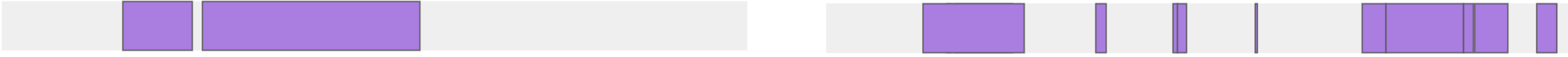}
    \end{minipage}  }    \\ \cline{1-9} 
    
& & & & & & & \cellcolor[HTML]{525252}{\textcolor{white}{V1}} & \cellcolor[HTML]{525252}{\textcolor{white}{V2}} \\
8 & 30m & 28m & 10m 47s & 7m 2s & 5 (5d) & 13(13d, 1t) &  \multicolumn{2}{|c|}{\begin{minipage}{.45\textwidth}
      \includegraphics[width=\linewidth, height=3.5mm]{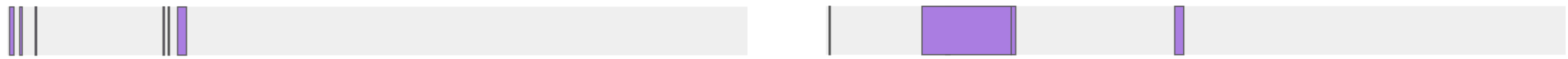}
    \end{minipage}  } \\
10 & 30m & 30m & 10m 51s & 7m 54s & 4 (4d) & 4 (3d, 1t) & \multicolumn{2}{|c|}{\begin{minipage}{.45\textwidth}
      \includegraphics[width=\linewidth, height=3.5mm]{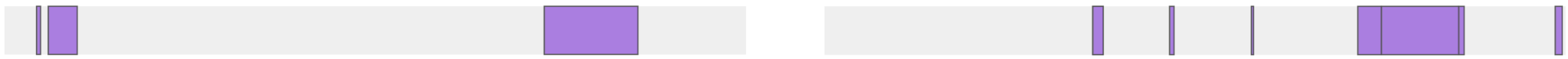}
    \end{minipage}  }   \\
11 & 30m & 22m & 8m 37s & 10m 45s & 21 (21d) & 4 (4d) & \multicolumn{2}{|c|}{\begin{minipage}{.45\textwidth}
      \includegraphics[width=\linewidth, height=3.5mm]{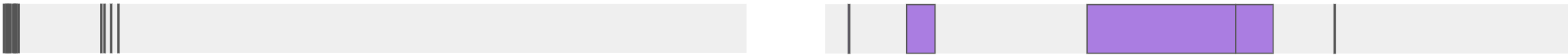}
    \end{minipage}  }  \\    
12 & 30m & 18m & 11m 4s & 7m 51s & 15 (15d) & 5 (5d) & \multicolumn{2}{|c|}{\begin{minipage}{.45\textwidth}
      \includegraphics[width=\linewidth, height=3.5mm]{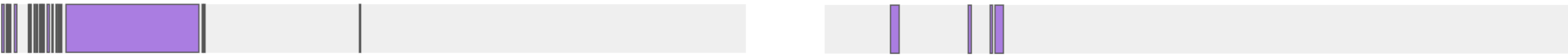}
    \end{minipage}  }    \\
14 & 30m & 30m & 10m 44s & 7m 30s & 8 (8d) & 12 (12d) & \multicolumn{2}{|c|}{\begin{minipage}{.45\textwidth}
      \includegraphics[width=\linewidth, height=3.5mm]{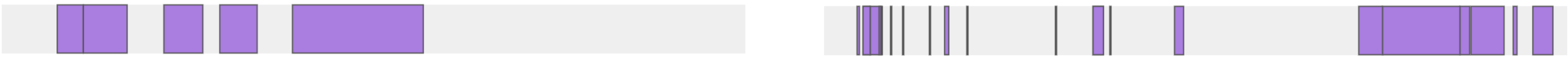}
    \end{minipage}  }    \\
17 & 30m & 25m & 11m 24s & 9m 26s & 9 (4d, 4i, 1a) & 27 (24d, 3t) & \multicolumn{2}{|c|}{\begin{minipage}{.45\textwidth}
      \includegraphics[width=\linewidth, height=3.5mm]{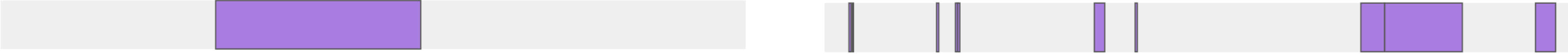}
    \end{minipage}  } \\\hline    
\end{tabular}
\end{table*}

\subsection{Results}
Overall, participants rated using ~\sysname{} to edit videos as requiring significantly less effort ($\mu$=4.58, $\sigma$=1.51 vs. $\mu$=2.17, $\sigma$=1.11; $Z$=2.96, $p$<0.01), frustration ($\mu$=3.58, $\sigma$=2.11 vs. $\mu$=2.08, $\sigma$=1.31; $Z$=2.39, $p$<0.05), mental demand ($\mu$=3.17, $\sigma$=1.80 vs. $\mu$=2.00, $\sigma$=0.95; $Z$=2.03, $p$<0.05), and temporal demand ($\mu$=4.5, $\sigma$=1.93 vs. $\mu$=2.58, $\sigma$=1.16; $Z$=2.54, $p$<0.05) compared to their own editing tools that they were experienced with (Figure~\ref{fig:NASA-TLX}). Perceived performance and physical demand were not significantly decreased for ~\sysname{}, and all significance testing was performed with the Wilcoxon Signed Rank test. 
All participants stated they would like to use ~\sysname{} in the future for reviewing and editing their videos.
% Using ~\sysname{} participants were able to edit their video more efficiently, such that their final output video contained more edits with 
% ~\sysname{} than with their typical video editing tools. 
% When reflecting on their final video, participants expressed they were more confident with their final result, and needed less assistance reviewing it compared to their typical editing tools. 

We report the statistics of the videos edited by participants in ~\autoref{tab:video_stat}. While 30 minutes were given for each editing session, six participants using ~\sysname{} finished the task early. Due to the limited time, ten participants using their own editing tools did not edit the later part of the footage (P4, P8, P9, P11-17). 
% The compression rates (final video length / original video length) of the baseline and \sysname{} conditions were not significantly different.
The Video Timeline column in \autoref{tab:video_stat} shows the edited time segments over the timeline of the videos. 
{As participants using their own tools often did not reach the second half of the video, the output videos in the baseline condition included notable errors in the latter half of the video such as leaving in dark scenes (V1), long pauses (V2), and repetitive actions (V2). However, across both conditions, short edits to the video timeline often introduced jump cuts ~\cite{ncsl, TTU} in the final output video.}
% Using \sysname{}, participants tended to edit a wider region on the timeline and edit more fine-grained segments.

\autoref{fig:sequence} summarizes how creators used ~\sysname{} by visualizing operation sequences relevant to navigation and editing.
%Jump
Overall, participants frequently jumped between different parts of the video using the headings, transcript lines, and words in the \textit{audio-visual script} (\autoref{fig:sequence}, light blue ``Text Jump'' cells). 
% Participants did not actively jump using the search; 
Four participants (P10, P14, P16, P17) used the search feature once (\autoref{fig:sequence}, blue ``Search Jump'' cells). 
%Deletion
Participants frequently deleted speech, pauses, and visual errors in the video (\autoref{fig:sequence}, yellow, orange and red ``deletion'' cells). 
Because \sysname{}'s ~\textit{audio-visual script} is aligned with the video timeline and contains descriptions of pauses and errors (\textit{e.g.,} duration, error type), five participants (P4, P8, P9, P16, P17) often subsequently deleted problematic segments only using text descriptions without actually playing the video.
%Other edits
In addition to deleting clips, participants tried to recover from pauses and visual errors by trimming or changing the speed; five participants trimmed pause segments (P8, P10, P15, P16, P17) and one participant changed the playback speed (P1).
% On the other hand, to correct the visual error only P1 changed speed.

% We 
% We report how ~\sysname{} improved participants' editing practices by addressing existing challenges (Section 3, \textbf{C1-C5}):
% across three stages of video editing and participants' assessment of the final video itself: 
% In addition, participants were able to make edits significantly more efficiently with ~\sysname{} than with their own editing tools and were more confident in their final video. 
% : effort ($\mu$=X,\sigma$=Y vs. $\mu$=X,$\sigma$=Y; $p$<$0.01$), frustration ($\mu$=X,\sigma$=Y vs. $\mu$=X,$\sigma$=Y; $p$<$0.05$), mental demand ($\mu$=X,\sigma$=Y vs. $\mu$=X,$\sigma$=Y; $p$<$0.05$), and temporal demand ($\mu$=X,\sigma$=Y vs. $\mu$=X,$\sigma$=Y; $p$<$0.05$) compared to their own editing tools that they were experienced with (Figure~\ref{fig:NASA-TLX}).

\subsubsection{Reviewing Videos and Identifying Errors to Edit. }
Participants rated ~\sysname{} as significantly more helpful for reviewing their video footage to identify errors compared to their existing editing tools ($\mu$=4.25, $\sigma$=2.22 vs. $\mu$=2.00, $\sigma$=1.04; $Z$=2.17, $p$<0.01). When reflecting on their final video, participants expressed that they were more confident with their final result ($\mu$=4.67, $\sigma$=1.37 vs. $\mu$=3.00, $\sigma$=1.41; $Z$=2.34, $p$<0.01), and needed less assistance reviewing it ($\mu$=5, $\sigma$=1.54 vs. $\mu$=2.75, $\sigma$=1.66; $Z$=2.31, $p$<0.01) compared to their typical process.
% editing tools. 

\ipstart{Text-based vs. Timeline-based Video Review} Using ~\sysname{}, participants primarily reviewed the video by reading the text of the audio-visual script and outline, while with their own video editing tools participants primarily reviewed the video by playing the video.
For example, of 7 participants who reviewed the entire video before editing it with \sysname{}, five participants read the entire audio-visual script using a screen reader or braille display without playing the video (P4, P9, P10, P16, P17), and three read the outline to gain an overview of the video (P10, P11, P13). P10 did both. On the other hand, when using their baseline tools, all participants played the video from the beginning to identify points to edit. Participants expressed that reading the text of the audio-visual script or outline allowed them to skim the footage faster than the video alone. P16 who reviewed the 11-minute video with ~\sysname{} in 3 minutes remarked, \textit{``I've been using NVDA [screen reader] for so long that I can understand a very fast TTS [Text-To-Speech]. Because I read 1,075 words-per-minute reading the script instead of playing video saves so much time for me.'' } 

\ipstart{Gaining an Overview of Visual Content and Errors} In addition to providing options for faster review, participants reported that they used ~\sysname{}'s high-level description of visual scenes and errors to (1) form a mental picture of the visual content (\textit{e.g.}, connecting background sounds with the scene descriptions, or imagining what the scene contained), (2) plan what edits they would like to make later (\textit{e.g.}, get an overview of the parts of the video that they needed to edit), and (3) mentally bookmark their progress as they edited (\textit{e.g.}, using a scene title to remember they had left off editing). P16 remarked that the descriptions were particularly helpful for silences: \textit{``Even in silence, I know what is going on in this video! Reading these scene labels, I can construct mental imagery of what the footage looks like.''} P10 and P14 also used the \textit{inspect} feature in concert with the high-level descriptions of visual content and errors (\textit{e.g.}, to understand the content of a pause, and to access objects at the beginning of a scene). 

\ipstart{Identifying Opportunities to Edit Video Footage} Participants considered ~\sysname{}'s visual errors in making decisions for visual editing, while they edited audio errors (\textit{e.g.}, pauses, and repeated words) with both systems. 
Using ~\sysname{}, all participants reviewed the visual errors in the video, and 11 of the 12 participants ~\sysname{} edited a visual error (\textit{e.g.}, by deleting, speeding up, or trimming the error). When evaluating visual errors, participants read the error along with the speech associated with the error to decide whether to delete it or not. For example, when assessing a visual error that overlapped with an important sentence in the speech that would harm the meaning of the speech if deleted, participants left the footage intact. On the other hand, if participants could make a natural edit (\textit{e.g.}, cutting out an unnecessary sentence, or trimming the length of the error) they would cut it out. To edit the errors detected by ~\sysname{}, 11 participants deleted the entire segment of the error, whereas one participant changed the playback of the error segment leaving some part of it. P13 stated, \textit{``If I just get rid of the error, it might result in a jumpcut or leave a too small gap between the sentences which is unnatural.''}
Participants expressed that getting informed of the visual errors made them more confident in their edits, but P4, P9, P11, and P12 noted they would like severity information about the error to inform quality vs. content trade-offs. P12 noted \textit{``It says bad lighting, but what I want to know is how bad so that I can make a decision whether to keep it, fix it, or remove it.''} 

With both systems, participants edited out irrelevant footage and audio errors (\textit{e.g}, pauses, repeated words). With ~\sysname{}, participants made edits at word level or line level (a sentence, a long phrase, or a pause) and sometimes removed multiple lines at once when they decided not to keep a big chunk of the scene that they did not find interesting. Using their own editing tools, all participants made edits to remove filler words or pauses between speeches, and some participants similarly deleted uninteresting content.

\subsubsection{Navigating and Applying Edits.} 

While participants found AVscript to be beneficial for high-level navigation and editing operations (\textit{e.g.}, by scenes, lines, words, long pauses) and non-linear navigation, the current version lacked the fine-grained navigation and editing provided by their typical video editors that enables participants to edit fine-grained audio (\textit{e.g.}, short pauses). As participants found ~\sysname{} to be helpful for some navigation tasks more than others, participants did not rate ~\sysname{} to be significantly more helpful for their existing tools for navigation ($\mu$=2.5,$\sigma$=2.11 vs. $\mu$=1.3,$\sigma$=0.78; $Z$=1.63, $p$>0.05) or applying edits ($\mu$=2.58,$\sigma$=1.73 vs. $\mu$=1.83,$\sigma$=1.19; $Z$=0.99, $p$>0.05). 

\ipstart{Coarse-Grained Navigation} Using ~\sysname{}'s audio visual script, all participants efficiently navigated the video content by moving the cursor in the transcript both line-by-line (up/down arrow keys) and word-by-word (alt/option + right/left arrow keys). P12 and P16 also jumped to the next scene in the audio-visual script by pressing the `H' key in the screen reader's browser mode (used to navigate to the next heading element). As participants edited the video, 7 participants also used the outline pane to quickly navigate to a scene or an error suggestion. In contrast, using their typical video editors' timelines all participants navigated by skipping ahead in a fixed time or frame interval (\textit{e.g.}, skip ahead 5 seconds) rather than by content (\textit{e.g.}, sentence, word, pause, error or scene). Participants then needed to iterate multiple times to find the relevant cut point, as described by P11: \textit{``To delete one word, I have to navigate so many times to precisely set the start and end of what I want to cut out. So I often create a small loop around the target just for editing.''} Four participants also scrubbed backward or forwards to navigate to a near word or pause target (P8, P10, P11, P14) despite its disadvantages: \textit{`The scrubbing audio makes no sense to me, but it can still be used to detect pauses''} (P11). P10 and P11 also used the tab key in Reaper to jump to the next audio peak to locate the end of long silences.

\ipstart{Fine-Grained Navigation} While ~\sysname{} makes editing words or pauses convenient, participants asked that in the future ~\sysname{} also include frame- and interval-level navigation to facilitate fine-grained adjustments to the cursor placement, especially when speech is not present. In addition, as the system limited the pauses displayed to screen reader users to 3s long to optimize skimming the audio-visual script, participants expressed that they wanted a mode for fine-grained edits that would display small pauses.

\ipstart{Non-linear navigation}
Participants also used ~\sysname{} to efficiently navigate the video non-linearly, using the outline to navigate to an error they would like to edit, then moving their cursors back to play from a few lines prior to figuring out where to make the edit by considering the audio content and the visual error together (P4, P9). 
% \amy{N} participants used the outline pane to quickly navigate to a new scene or an error suggestion \amy{(PX, PY, PZ)}. After skipping the video using the outline, two participants moved their cursors and played from a few lines prior to the point where they jumped (P4, P9). 
Four participants used the search pane to find and skip to a specific part in the script (P10, P14, P16, P17). P10 exclaimed: \textit{``This search feature is revolutionary! I can search not just for text, but an object or even pauses so easily.''} Yet, participants who never used the search feature to navigate the video speculated that searching for visual content would be more useful for their own videos. P9 noted \textit{``I didn't know what to search for as I didn't film this video. If I use it (\sysname{}) for my own video, I will definitely find it useful.''} 

\ipstart{Applying edits} The ability to apply edits with ~\sysname{} was limited to high-level edits of the video footage itself. With their own editing tools, participants additionally applied effects to improve the audio or visual quality of the footage, including: applying a high pass filter to remove background noise and heavy breaths (P13), inserting music and adjusted its volume so that the original audio is louder than the music (P17), adding an intro and credit to the footage by inserting a black image with white text (P17). After making a cut in the video, P15 and P17 used a transition effect to avoid the abrupt jump in the audio or visual. 
% P17 mentioned \textit{``I usually stick to one animation (blur through black) because my sighted friend said it is the safest one to use in most cases. It's hard to visualize the effect just by reading the animation descriptions.''}
With When making edits, 2 participants often referred to a help menu, or a self-created list of hotkeys and commands to remember the keys they should use (P4, P8). 
Participants who didn't use the built-in video player of the editing tool read the timestamps from the player and then passed them into the command line (P4 using FFmpeg), or to the input field of the tool (P16 using VideoReDo). Both P4 and P16 noted the inconvenience of switching between two different interfaces. P16 said ``\textit{Because the video player and VideoReDo use different time formats, I cannot directly copy and paste. When I manually read and type them, I sometimes make typos and this makes very confusing results.}'' P4 also mentioned ``\textit{While the script-based editing is very accessible, I have to run the command after each edit to check the results. If it's a long video, I have to wait for a long time for the video to be processed.}''

\section{Exploratory Case Studies}\label{exploratory-study}
The comparison study demonstrated that BLV creators were able to use \sysname{} to understand and edit videos.
% Through the comparison study, we could learn that BLV creators can understand the concept of \sysname{} and actively use the main features to optimize given videos. 
To learn how BLV creators would use ~\sysname{} to edit their own videos, we conducted an exploratory study with 3 BLV creators (P14, P18, P19 in Table~\ref{tab:participants}) where the creators edited their own footage.
% using ~\sysname{} to examine: How would BLV creators use ~\sysname{} to edit their own videos? 
% Our goal was to assess the potential of ~\sysname{} in a more realistic scenario. 

\subsection{Method}
\begin{figure}
  \centering
  \includegraphics[width=0.45\textwidth]{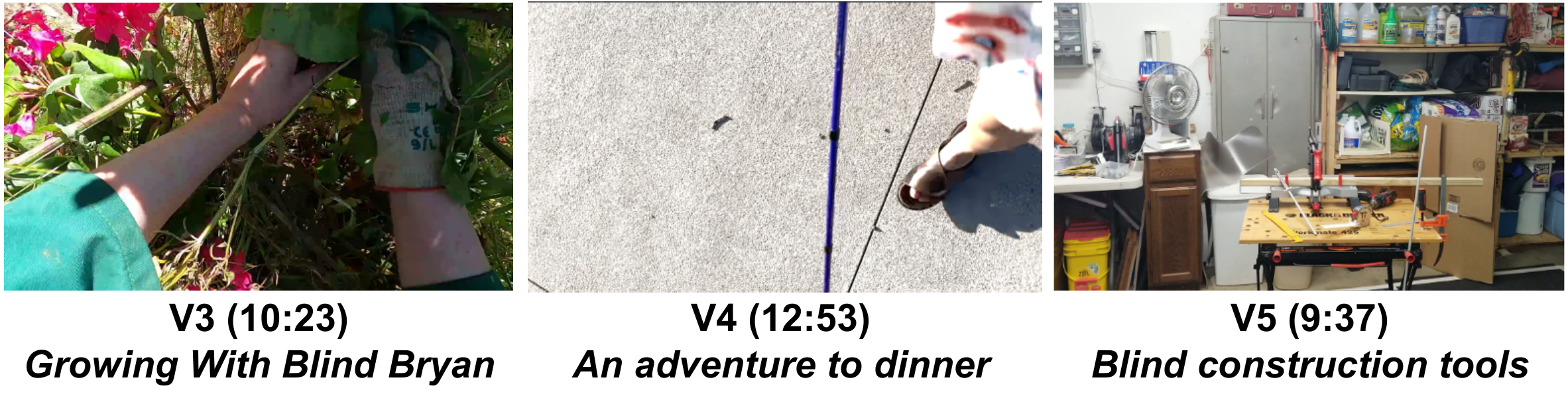}
  \caption{Three videos filmed by BLV creators for exploratory case studies.}
  % (Section \ref{exploratory-study}).}
  \label{fig:exp-videos}
  \Description{The thumbnails of the three videos used in the exploratory case studies. In the first image, the hands of a gardener can be seen. In the second image, a woman is walking with a purple cane. In the third image, the inside of a garage is shown.}
\end{figure}

We recruited 3 video creators with visual impairments who used screen readers to access their devices using mailing lists and social media (P14, P18, P19). P14 also participated in the comparison study.
% (P14). 
% Similar to the comparison study~\ref{comparison-methods}, we recruited participants who had a visual impairment and used a screen reader to access their device. 
All three participants created and uploaded videos to their YouTube channels on a regular basis, and two of the three participants had not edited videos before. Before the study, we collected footage from each participant that they had filmed themselves (\autoref{fig:exp-videos}).If participants provided multiple clips we concatenated them in order of time filmed.  
% and created a single video by concatenating clips in order of time filmed if participants wanted to use multiple clips (\autoref{fig:exp-videos}). 
% The types of the videos were: 1) Planting tutorial (V3) 2) Vlog (V4) 3) Construction tools review (V5). None of the collected footage was screen-recording. 
% If there were several clips that participants wanted to use for the editing study, we concatenated the clips in the order of the time each was filmed.
During a 120-minute remote study session, we asked participants background questions, provided a tutorial of ~\sysname{}, invited participants to edit their own footage with ~\sysname{}, and asked participants semi-structured interview questions about their experience (see Supplemental Material). We compensated participants \$80 via Amazon Gift Card for filming their footage and participating in the study.
% The 
% We conducted a 120-minute remote study using Zoom~\footnote{https://zoom.us/}. We first asked participants demographic and background questions about their prior video editing experience.
% We then gave a 20-minute tutorial on the ~\sysname{} interface and the participants practiced editing V0 to familiarize themselves with the system features.
% Participants then edited their own footage using ~\sysname{}. After the editing session, we ran semi-structured interviews to gain a deeper understanding of the strengths and weaknesses of ~\sysname{} and the decisions about specific referencing strategies participants employed or interesting usage patterns. We compensated participants with a \$80 Amazon Gift Card for their participation and the time filming the footage.

\subsection{Three Vignettes: How BLV Creators Use ~\sysname{} in Context}
\subsubsection{V3: Growing with \andrew{}}
% : Reviewing footage with ~\sysname{}}
\andrew{} (P18) regularly posts videos to demonstrate how nature is accessible on his YouTube channel. To film his planting demonstration video (V3), he strapped a camera to his chest or forehead to use both hands freely and filmed four clips over four different days. Because the first two clips were filmed more than a month ago, \andrew{} quickly reviewed the footage by skimming through ~\sysname{}'s script. He mentioned \textit{``I usually make videos comparing how a plant changed after several months. Using [\sysname{}], I don't have to watch all the clips again. I can save so much time reviewing and remembering what I filmed!''} 
With ~\sysname{}, \andrew{} first used the ~\textit{outline} to jump to the start of each scene (clip), then deleted the first few lines of the script where he mentioned the date it was filmed. 
When he noticed that he pointed at a plant to describe it in the video, he used the \textit{inspect} feature to make sure the plant was in the frame.
\andrew{} described that with ~\sysname{} that he can be more independent in making videos, which will enable him to create videos more quickly. He explained that he typically required sighted reviewers: \textit{``Because it is so difficult to make sure everything I mention is in the picture, I usually film the same content with several takes, and ask sighted friends to ask which one is the most appropriate.''} 
% no longer have to film multiple takes, don't need to wait to get assistance from someone else - can work according to his schedule
% audio description

\subsubsection{V4: An Adventure to Dinner}
% : Recovering from mistakes using ~\sysname{}}
\rachel{} (P19) is a content creator who makes a wide range of different media: podcasts, interviews, live streams, Vlogs, and tech demos. While she is an experienced audio editor, she has never tried editing a video due to a steep learning curve. For the study, she filmed a Vlog on her way to dinner (V4). 
\rachel{} mentioned that ~\sysname{} is easy to learn and use with a screen reader: \textit{``Absolutely fantastic, I have never been able to edit videos before, but after 15 minutes of learning how to use this, I can edit my video. It's a giant leap forward.''}
While editing, \rachel{} found ~\sysname{}'s search feature useful because she still remembered most content of the footage that she filmed two days ago: \textit{``When I was walking on the street, I met a family and chatted with them for a while. To jump and edit that part, I tried searching for `boy' or `person'.''} She enjoyed having the option to search for the visual content of the footage, as she might forget the exact word she said, but still remember what was visible in the frame.
\rachel{} also noticed that ~\sysname{} had errors in the speech-aligned transcript and scene description. When she read one of the scene labels, she said \textit{``Oh it says I'm holding a purple leash! That is my purple cane. I guess this is created by AI?''}

Overall, \rachel{} mentioned that she feels more confident showing her video to more people after fixing the visual errors detected by ~\sysname{}. As a creator without light perception, \rachel{} has often experienced filming videos with bad lighting (\textit{e.g.,} turning the light off, or facing back to the sunlight). She noted \textit{``[\sysname{}] is also guiding me on how to film with fewer visual errors.''}

\subsubsection{V5: Blind Construction Tools}
% : \mina{should come up with the title} }
% being more self
\lewis{} creates workout videos, product reviews, and tips for people with visual impairments. He often shares his videos on social media or participates in workout video contests. To film a video on construction tools (V5), P14 set up a camera with a tripod and used TalkBack to guide him with the filming position (e.g., TalkBack giving directions ``Face detected - upper right''). 
% He has experienced video editing using Reaper. 
To quickly skim through his footage, \lewis{} pressed and held the down arrow key to mimic the scrubbing feature of Reaper (his typical video editor). He described that the lines helped him navigate efficiently: \textit{``I don't have to read the entire line to check where I am (the cursor is) in the video. Just listening to the first word or first syllable is enough.''}
When \lewis{} reached a part of the video that ~\sysname{} detected as \textit{blurry} he mentioned: \textit{``Oh this is not a bad thing here, I had to walk quickly, and it's probably because of that.''}
\lewis{} also used the ~\textit{inspect} feature to choose an editing point. To find the first few seconds where he started the recording and was not on the frame, he continuously clicked ~\textit{inspect} to find the exact timestamp where he appeared, then trimmed the video up to that point. He noted: \textit{``The script does tell when the word begins and ends, but it doesn't tell when an action begins and ends.''} \lewis{} also reported speech recognition errors: \textit{``I mumbled something here, but it wasn't caught in the transcript. Maybe because of the radio music. It is difficult to edit that part out when I don't see it on the transcript.''}

Using ~\sysname{}, \lewis{} anticipated that collaboration with sighted reviewers will be easier because he can show them only the errors detected by the system instead of asking them to review the entire footage. He also noted that he wanted to create different content and styles of videos with the help of ~\sysname{}: \textit{``In the past, I always used a tripod to avoid camera shakes. Now that I can check whether my footage is shaky, I want to try carrying around my camera.''}

\subsubsection{Reflection on Three Vignettes}
% using their own videos, limitations, future workflow 
All three BLV creators used AVscript to speed up video editing steps (\textit{e.g.,} \rachel{} browsing the video for a specific scene), or locate visual errors or actions (\textit{e.g.,} \lewis{} noticing a camera blur) which was not possible prior to using ~\sysname{}. 
Creators also reported the limitations of ~\sysname{}: (1) errors in the speech-aligned transcript and scene description, (2) lack of detailed information on visual content such as motion details or object colors, especially for clips without much narration.
Overall, all three creators wanted to use ~\sysname{} in the future to be more creative with the content and styles of videos.

\section{Discussion and Future Work}

{In this section, we reflect on our findings from the 
design, development, and evaluation of \sysname.
% comparison studies (Section \ref{comparison-study}) and the exploratory case studies (Section \ref{exploratory-study}). 
We also present future directions for research exploring accessible authoring tools.}

\begin{revisedenv}\ipstart{Navigating Videos based on Visual Content}
% connecting back to formative 
% more clear distinct btw prior work, our work, and future work
Our formative study revealed that BLV video creators' current tools did not enable access to visual content in their video footage (\textbf{C1.} Recognizing visual content in a video). 
To address this challenge, \sysname{} provides access to visual content including: a summary of key visual moments via \textit{scene descriptions}, a list of low-level objects via \textit{inspect mode}, and on-demand access to visuals of interest via \textit{search}. 
While creators using \sysname{} occasionally listened to the video and scene descriptions linearly (similar to how BLV audiences currently listen to audio descriptions that describe visual content in a video alongside the video narration~\cite{pavel2020rescribe,li2021non}), creators also used scene descriptions for new use cases including skimming the outline of scene descriptions to gain an overview of visual content, and clicking on scene descriptions to navigate to video scenes (similar to how sighted video creators use video keyframes to navigate with timeline-based video editing tools~\cite{finalcut, premiere}).
Scene description-based navigation helped address an existing challenge for video creators (\textbf{C4}. Non-linear browsing and skimming of videos), and future work may explore extending this navigation approach to video consumers.
However, state-of-the-art scene descriptions still include errors. In our studies, BLV creators editing their own footage were able to recognize and recover from errors that mismatched their expectations (\textit{e.g.}, ``leash'' vs. ``cane'' in a walking video), but creators editing unfamiliar footage missed notable errors (a pile of laundry described as ``animal on bed''). 
Future improvements to scene description accuracy could help \sysname{} better support BLV creators using unfamiliar footage (\textit{e.g.}, when adding stock video b-roll).
While creators used \sysname's low-level inspect mode less often than the high-level scene descriptions, one creator used inspect mode to achieve fine-grained navigation to a visual scene boundary, similar to the fine-grained navigation to audio pause boundaries that BLV creators currently perform via audio. 
Future work may explore how navigation practices change with long-term use of \sysname{} for video editing, and how to further facilitate fine-grained visual navigation.
\end{revisedenv}

\begin{revisedenv}\ipstart{Editing based on Visual Error Suggestions}
To address the challenge of assessing the visual quality of a video (\textbf{C2}), \sysname{} informs users of potential edit points (blur, bad lighting, camera motion, and audio pause). 
% In the comparison study, participants edited out more visual errors when using \sysname{} than using their own tools (ref to new evaluation). 
% In the comparison study, participants reported significantly more confidence in the final video quality when using \sysname{} than using their own tools. Yet, the remaining question is how users notice and handle incorrect error suggestions from \sysname{}. 
% Participants reported that \sysname's edit point suggestions improved participant confidence in the final video.
Participants frequently used the visual errors provided by \sysname{} to remove distracting and low-quality visuals from the video (\textit{e.g.}, camera shakes, dark lighting), and participants reported that edit point suggestions improved their confidence in their final video. 
While participants occasionally noticed errors in visual content descriptions, none of the participants expressed skepticism toward visual quality predictions. 
However, participants asked for information about the severity of the predicted visual errors to help them weigh the content and quality of a clip before removing it.
% could better consider the trade-offs in removing a clip between the content and quality of a given clip during editing.
In the future, \sysname{} will provide confidence scores and severity levels for predicted visual errors to better support BLV creators in making editing decisions. 
In addition, describing errors in more detail and explaining potential causes (\textit{e.g.,}``\textit{Blur -- out of focus, possibly due to the camera being too close to the target object}'') could help novice video editors understand errors and film better footage. Finally, \sysname{} could recognize other types of visual errors in the future, such as \textit{composition errors}~\cite{adams2013qualitative} and \textit{jump cuts}~\cite{TTU} which we observed in the videos edited by BLV creators.
\end{revisedenv}

\begin{revisedenv}
\ipstart{Text-based Video Editing for BLV Creators}
Prior research on text-based editing primarily focused on sighted video authors~\cite{truong2016quickcut, huber2019bscript, Berthouzoz2012}. We explored using text-based editing to help BLV creators navigate videos efficiently (\textbf{C3}). Our studies revealed benefits of using text-based editing that echo findings in prior work (\textit{e.g.}, lower mental load than timeline-based interface \cite{truong2016quickcut, huber2019bscript}), and demonstrated unique benefits for BLV creators (\textit{e.g.}, better screen reader compatibility than timeline-based interfaces, and access to rapid screen reader text-to-speech for quick skimming and editing).
Still, creators mentioned that timeline-based video editing interfaces enable granular access to the audio track without word-level constraints (\textit{e.g.,} editing out background noise which is not captured in the transcript). 
% Yet, screen reader users can only jump in fixed time units and cannot use a mouse for scrubbing the timeline. 
% To design an efficient timeline-based video control, future research can create an audio thumbnail that provides a short audio snapshot of scenes for navigation. 
In the future, we plan to integrate timeline-based editing into ~\sysname{} to enable creators to use the \textit{audio-visual script} for coarse navigation and the timeline for fine-grained navigation.
\end{revisedenv}
% One of the interesting findings was that BLV users had a different mental model of editing video via text compared to sighted users who have access to a video player while editing the script. For instance, some study participants mentioned that the task felt like an essay-editing task, making them focus too much on grammar instead of considering the natural flow of colloquial words. }

% \revised{\ipstart{Supporting Broader types of Visual Error}
% move to the visual error suggestion section, what part of guideline we missed, jumpcuts
% }

\ipstart{Supporting New Editing Tasks}
While our system explored deleting or speeding up video segments --- core tasks in video production --- future can explore how to support BLV creators in making visual edits such as composing title slides or adding visual effects. For example, an editing system could describe the impact of an applied effect on the visual content in the video (\textit{e.g.}, ``the vignette effect now covers the hands'') using techniques from prior work in BLV visual design authoring~\cite{peng2022diffscriber} and computer vision approaches for captioning differences between pairs of similar images~\cite{jhamtani2018learning}. 
Recent strides in prompt-driven text generation~\cite{brown2020language}, image generation~\cite{ramesh2022hierarchical, saharia2022photorealistic}, and image editing~\cite{nichol2021glide} suggest that prompt-driven video editing (\textit{e.g.}, make this clip moody) may be possible in the future~\cite{hong2022cogvideo}. Future research is needed to help BLV creators evaluate their results with such tools. 
In addition, ~\sysname{} considers single-track videos as the format common in BLV creators' videos today. However, in the future, we could explore approaches to help creators enhance their videos with b-roll (\textit{e.g.}, by helping creators find and insert their own footage using text~\cite{truong2016quickcut} or suggesting opportunities for adding online b-roll~\cite{huber2019bscript}).

% \subsection{Supporting different stages of video creation process}
\ipstart{Supporting New Stages of Video Production}
Our formative study suggested that BLV creators currently use creative but effort-intensive filming strategies (\textit{e.g.}, describing visual content) and editing strategies (\textit{e.g.}, navigating video footage only linearly) to produce and share their videos to broad audiences.
% creative strategies in both filming and editing to achieve their final result. 
As ~\sysname{} enabled BLV creators to edit videos more efficiently, with less mental demand, and more confidence in their end result, BLV creators mentioned it would change their filming practice by capturing additional desired footage.
Future work may explore how improved access to video editing will impact filming practices, and further improve the filming process by providing additional information about the visual content and errors, as provided in our system, at capture time. Similar to prior work in supporting BLV photographers, future systems could information about framing the shot~\cite{adams2016blind} paired with the presence and severity of potential visual errors. When BLV creators move as they film videos like Vlogs and tutorials, approaches to alert creators of potential errors may distract them (similar to the demand of describing visual content). Thus, future work could explore enabling BLV creators to capture with a wide field of view at capture time (\textit{e.g.}, 360 or 180-degree video) and edit the video to produce a smooth normal field of view footage that captures the content of interest~\cite{su2016pano2vid}. 
Finally, our work points to solutions in the video publishing process including improving the acceptance of sighted audiences to visual errors, and platform-supported funding for BLV creators seeking to hire sighted reviewers. 

% \subsection{Adapting ~\sysname{} for Broader Contexts}
% \sysname{} informs BLV creators about visual content and visual errors by describing a scene or notifying errors. Yet, our study participants raised the issue that the amount of and the modality of the information cannot be customized 

% multimodal feedback,
% granularity, 

% \subsection{Collaboration Support}
% From the study, participants metn
% comments

% % describing visual errors/

\ipstart{Beyond Manual Text-based Editing}
We designed ~\sysname{} to use text as BLV video creators we interviewed were highly proficient at using screen readers. Text enabled creators to use their screen reader experience to review and navigate video at high speeds. 
We are currently exploring multimodal approaches for editing videos by combining a screen reader and voice input together to facilitate fast and low-burden navigation and editing (\textit{e.g.}, ``\textit{jump to 5 minutes}'', ``\textit{delete this scene}'').   The visualization community has demonstrated ample applications that support multimodal data exploration with touch and speech~(\textit{e.g.,}~\cite{Srinivasan2018Orko, Srinivasan2020DataBreeze, Srinivasan2020InChorus, kim2021data}). In similar vein, we plan to build on work in tactile displays~\cite{zhang2020automatic} to surface the visual content in the video. While consuming video with tactile displays may be challenging, editing video may benefit from providing creators access to slow frame-by-frame content (\textit{e.g.}, to assess when a person moves out of the frame) and waveform visualizations. 
\section{Conclusion}
In this work, we designed and developed ~\sysname{}, an accessible text-based video editing tool that enables BLV creators to edit videos using a text script that describes the visual content and visual errors in the footage. The design was informed by a formative study consisting of YouTube video analysis and interviews with BLV creators. The comparison study (N=12) showed that \sysname{} significantly reduces the mental demand of BLV creators when compared to their own video editing tools. In the exploratory case study (N=3) we also explored how BLV creators edit their own videos using \sysname{}. We hope our research catalyzes future work on improving the accessibility of media authoring.
% Our comparison study revealed that BLV creators quickly learned the core concept of \sysname{} and leveraged them to optimize raw video footage. 

%%
%% The acknowledgments section is defined using the "acks" environment
%% (and NOT an unnumbered section). This ensures the proper
%% identification of the section in the article metadata, and the
%% consistent spelling of the heading.
\begin{acks}
This work was in part supported by NAVER AI Lab as a research internship and the University of California, Los Angeles.
\end{acks}

%%
%% The next two lines define the bibliography style to be used, and
%% the bibliography file.
%\typeout{}
\bibliographystyle{ACM-Reference-Format}
\bibliography{sample-bib}

\clearpage
\onecolumn
\appendix
%\clearpage

\section{STUDY PARTICIPANTS DEMOGRAPHICS}

\begin{table*}[htp]
\aptLtoXcmd{}{%
    \small\sffamily
			\def\arraystretch{1.2}
		    \setlength{\tabcolsep}{0.3em}
}
		    \centering
\caption{Study Participants (P1-P8: Formative study participants, P1, P4, P8-P17: Controlled study participants, P14, P18, P19: Exploratory study participants. Three participants marked with {$^{*}$} participated in both formative study and controlled study (P1, P4, P8), and one participant with ~\dag~ participated in both controlled study and exploratory study (P14). All participants are screen reader users.)}
  \label{tab:participants}
%\centering
  \begin{tabular}{lccccccc}
    \toprule 
    PID & Age & Gender & Visual impairment & Onset & Video editing tool & Content type & Experience (yr.)\\ 
    \midrule
    P1{$^{*}$} & 27 & M & Legally blind & Acquired & Microsoft Photos & User interviews & 4 \\ %Vaibhav
    P2 & 23 & M & Totally blind & Acquired & VirutalDub 2 & Sports videos, Product reviews & 5\\ %Ross
    P3 & 22 & F & Legally blind & Congenital & Final Cut Pro & Live streams, Presentations  & 1\\ %Taylor
    P4{$^{*}$} & 35 & M & Low vision & Acquired & Python \& FFmpeg & Art demonstrations, Tutorials & 7 \\ %William
    P5 & 28 & F & Legally blind & Congenital & iMovie & Video podcasts & 11\\ %Casey
    P6 & 24 & M & Low vision & Acquired & Final Cut Pro & Short-form videos & 4\\ %Bruce
    P7 & 41 & M & Legally blind & Congenital & iMovie (mobile) & Vlogs & 2\\ %Jon
    P8{$^{*}$} & 41 & M & Legally blind & Acquired & Final Cut Pro & Short film & 20\\ %Juan
    P9 & 40 & F & Totally blind & Congenital & Microsoft Photos & Accessibility videos & 1 \\ %Anna
    P10 & 54 & M & Totally blind & Congenital & Reaper & Podcasts, Music videos & 1\\ %Christopher
    P11 & 31 & F & Legally blind & Acquired & Reaper & Fashion videos, Accessibility videos & 8\\ %Marche
    P12 & 30 & M & Totally blind & Congenital & Reaper & Twitch streams, Short-form videos & 3\\ %Florian
    P13\dag & 58 & M & Legally blind & Acquired & Reaper & \makecell{Workout videos, product reviews,\\Accessibility videos} & 2\\ %Lewis
    P14 & 40 & M & Legally blind & Congenital & Reaper & Tech demonstrations & 1\\ %Rod
    P15 & 29 & F & Totally blind & Acquired & Windows Movie Maker & \makecell{Accessibility videos,\\Video editing tutorials} & 5\\ %Karla
    P16 & 31 & M & Totally blind & Congenital & VideoReDo & Conference talks & 6\\ %Stefan
    P17 & 30 & F & Totally blind & Acquired & Windows Movie Maker & Video podcasts, Short-form video & 3\\ %Jaelene
    P18 & 63 & M & Totally blind & Congenital & None & Planting tutorials & 5\\ %Andrew
    P19 & 29 & F & Totally blind & Acquired & None & Vlogs, Video podcasts, Live streams & 0\\ %Rachel
   \bottomrule   
\end{tabular}
\end{table*}

\section{EVALUATION STUDY VIDEO DATASET}
\begin{table}[htp]
\aptLtoXcmd{}{%
    \small\sffamily
			\def\arraystretch{1.2}
		    \setlength{\tabcolsep}{0.3em}
}
		    \centering
\caption{Videos used in the evaluation study. }
\label{tab:videos}
\begin{tabular}{ccccc}
\toprule
\multicolumn{1}{l}{\textbf{Video ID}} & \textbf{Title} & \multicolumn{1}{l}{\textbf{Duration (Original)}} & \textbf{Creator} & \multicolumn{1}{l}{\textbf{URL}} \\ 
\midrule
V0 & College Life...As A Blind Girl! & 3m 12s (9m 10s) & Rae Green & \cite{v0} \\ 
V1 & How Blind Mom Cooks & 11m 12s (20m 38s) & Ashley Nemeth & \cite{v1} \\ 
V2 & Day In The Life Blind Mom & 11m 5s (20m 16s) & Ashley Nemeth & \cite{v2} \\ \hline
V3 & \makecell{Growing With Blind Bryan: \\ New border, Rampant Runner and a Juicy Peach Tree} & 10m 2s & P14 & None \\ 
V4 & An adventure to dinner: Demonstrating O\&M techniques & 12m 57s & P18 & None \\ 
V5 & Blind construction tools & 9m 37s & P19 & None \\ 
\bottomrule  
\end{tabular}
\end{table}

\section{FORMATIVE STUDY VIDEO DATASET}
To collect videos demonstrating how BLV creators edit videos, we first searched YouTube for all combinations of a set of vision-related keywords (blind, low vision, visual impairment, screen reader), and a set of video editing keywords (editing videos, making videos, creating videos), following prior work~\cite{li2021non, li2022feels}.
For each search phrase, we included all unique videos that had a title related to vision and video editing and stopped the search when the results of an entire search page were irrelevant.
We then filtered out videos that did not cover video editing (1 filtered) or had poor audio or video quality (3 filtered). Our final dataset contained 24 videos (V1-V24) uploaded before October 12, 2021.
% that averaged \amy{XX.XX} seconds long (min = 5:35, max = 30:56). 
The videos contained overviews of the video production process (2 videos) and tutorials of video editing software (22 videos). For the full list of videos, see Supplemental Material.
% \amy{list the videos?}

\end{document}